\PassOptionsToPackage{table,xcdraw}{xcolor}
\documentclass[sigconf]{acmart} 
\usepackage[font=small]{caption}
\usepackage[skip=0pt]{subcaption}
\AtBeginDocument{%
  \providecommand\BibTeX{{%
    \normalfont B\kern-0.5em{\scshape i\kern-0.25em b}\kern-0.8em\TeX}}}

\begin{document}

\title[Compatibility of Fairness Metrics with EU Non-Discrimination Law: DP \& CDD]{Compatibility of Fairness Metrics with EU Non-Discrimination Laws: Demographic Parity \& Conditional Demographic Disparity}
\author{Lisa Koutsoviti Koumeri}
\authornote{The first three authors contributed equally to this research. Yousefi covered the legal analysis. Koutsoviti Koumeri and Legast conducted the experiments and the technical analysis.}
\email{lisa.koutsoviti@uhasselt.be}
\affiliation{
\institution{Hasselt University}
\country{Belgium}
}
\author{Magali Legast}
\email{magali.legast@uclouvain.be}
\orcid{0000-0003-4246-1158}
\affiliation{
\institution{Université Catholique de Louvain}
\country{Belgium}
}
\author{Yasaman Yousefi}
\email{yasaman.yousefi3@unibo.it}
\orcid{0000-0003-1483-2978}
\affiliation{
  \institution{University of Bologna/University of Luxembourg}
 \country{Luxembourg}
}

\author{Koen Vanhoof}
\orcid{0000-0001-7084-4223}
\affiliation{
\institution{Hasselt University}
\country{Belgium}
}

\author{Axel Legay}
\orcid{0000-0003-2287-8925}
\affiliation{
\institution{Université Catholique de Louvain}
\country{Belgium}
}

\author{Christoph Schommer}
\orcid{0000-0002-0308-7637}
\affiliation{
  \institution{University of Luxembourg}
 \country{Luxembourg}
}
    \begin{abstract} 
    Empirical evidence suggests that algorithmic decisions driven by Machine Learning (ML) techniques threaten to discriminate against legally protected groups or create new sources of unfairness. This work supports the contextual approach to fairness in EU non-discrimination legal framework and aims at assessing up to what point we can assure legal fairness through fairness metrics and under fairness constraints. For that, we analyze the legal notion of non-discrimination and differential treatment with the fairness definition Demographic Parity (DP) through Conditional Demographic Disparity (CDD). We train and compare different classifiers with fairness constraints to assess whether it is possible to reduce bias in the prediction while enabling the contextual approach to judicial interpretation practiced under EU non-discrimination laws. Our experimental results on three scenarios show that the in-processing bias mitigation algorithm leads to different performances in each of them. Our experiments and analysis suggest that AI-assisted decision-making can be fair from a legal perspective depending on the case at hand and the legal justification. These preliminary results encourage future work which will involve further case studies, metrics, and fairness notions.
    \end{abstract}
\settopmatter{printfolios=true} 
\maketitle
    \section{Introduction}\label{sec:intro}
    With the rapid development of modern technologies in every sector of society and our reliance on automation, many legal and ethical concerns have been raised regarding how fair algorithmic decisions can be, and how much we can trust such decisions \cite{kleinberg2018algorithmic} \cite{mitchell2021algorithmic}. 
    Nowadays, our future, success, and well-being can be decided by algorithms and through algorithmic predictions that are based on past events: whether one shall be trusted with a loan or not, whether one shall be held in prison or not \cite{angwin2016propublica}, if students shall pass or fail an exam they have not even taken \cite{kelly2021tale}, or who should get access to certain opportunities including job offers \cite{lambrecht_algorithmic_2018, datta_automated_2015}, housing \cite{housing}, or even medical treatments \cite{manrai2016genetic}.
    Algorithmic decision systems are perceived to be impartial and evaluate all individuals in the same way and supposedly prevent discrimination \cite{citron2014scored}.
    However, specially due to their reliance on historical data, algorithmic decisions are not bias-free, just like human decisions. The wrong-doings of the past can be reinforced into the alleged fair decisions of the future in such data-based applications \cite{wachter2021fairness}.
    Many solutions have been offered to try and reach the fairness we see fit for humans in machines, or to prevent human discrimination and bias by relying on automation \cite{hort2022survey} \cite{madaio2020co}. Yet, no optimal and consensual solution exists.
    Even if we managed to eliminate historical bias from these systems, new grounds of discrimination might still find their way in. Further, we can argue that evaluating all individuals the same way is not necessarily what is fair, for example in some contexts for historically underprivileged social groups or in the medical field when differences need to be taken into account \cite{manrai2016genetic}.
    
    Defining what fairness is and what constitutes discrimination is a hard task. This has led scholars to introduce several definitions of fairness in different fields of study. In European Union's (EU) legal system, fairness is generally approached through a non-discrimination framework. In the EU, everyone is equal before the law and discrimination based on several grounds is prohibited. 
    
    Many scholars have argued that fairness is not a measurable value that can be expressed mathematically, but a human value that depends on the circumstances and settings of events \cite{wachter2021fairness}.
    Others -mostly in computer science- have explored ways to evaluate different fairness concepts through mathematical definitions in order to bridge the gap between the immeasurable notion of fairness and the technical needs of Artificial Intelligence (AI) \cite{kamiran2013quantifying} \cite{verma2018fairness}. Most of the existing literature consider that we cannot capture the essence of fairness with a one-size-fits-all solution in computer science.

    This research adds on the existing literature regarding bias mitigation as well as helps fill the gap between law and computer science to optimize the legal notion of equality in a technical reality. We build on the work of both legal and Machine Learning (ML) scholars to provide a theoretical reflection and a practical experimentation to assess how mathematical evaluations of fairness can be used to mitigate discrimination identified as such by the law in algorithmic settings. Conversely, we assess how non-discrimination law could use some technical tools to detect discrimination.

    We address this question through a case-study on the canonical classification problem involving widely used real world datasets (COMPAS \cite{dressel2018compas}, Adult \cite{Dua2019adult} and Law \cite{WinNT}). We explore the use of Demographic Parity (DP) \cite{dwork2012fairness}, a statistical definition of fairness, both as a fairness metric and as a constraint in a bias mitigation algorithm, as well as Conditional Demographic Disparity (CDD) \cite{wachter2021fairness}, introduced by legal scholars. We assess their relevance from a ML point of view and following a legal-informatics methodology. The latter expands on the way law shapes technology, and the impact of technology on law and legal concepts \cite{biasiotti2008legal}. This methodology can be applied to interpret and align the legal concept of fairness to the emergence of new technical paradigms.

    This work includes six sections. Following the introduction, section \ref{sec:related_work} studies the existing literature on fairness and ML and the relevance of this literature with our work. Section \ref{sec:fairness} gives an analysis of the concept of fairness from legal and technical points of view for the sake of the multi-disciplinary approach of this paper. In section \ref{sec:exp_setup}, we report the setup of our experiment, describing the methodology, datasets and algorithm used. In section \ref{sec:results}, we present the experimental results along with their legal and technical interpretations. We then conclude our research in section \ref{sec:ccl} by emphasizing on the importance of the choice of fairness metrics in bias mitigation, as well as considering the impact of human bias in algorithmic discrimination.
    %
    \section{Related work} \label{sec:related_work}
    Scholars in different sectors have explored fairness and bias in ML over the past decades. 
    Bias can creep into systems through different ways \cite{mehrabi2021survey}. For example, the algorithm itself can introduce biases \cite{friedman1996bias}. In many cases, the
    historical data that is used to train AI models using ML techniques contains underlying biases that is learned by the system. Those biases can then be seen in discriminatory predictions against certain groups of people with protected characteristics. If those characteristics are recognized in the legal framework, those groups are protected by the law in case of direct or indirect discrimination.
    
    It is important to detect, measure and mitigate those biases, with the aim to obtain fairer decisions. This shall be done from ethical, legal and technical perspectives due to the relevance of the topic in all domains. Efforts in this direction include the introduction of fairness metrics and bias mitigation methods, as well as new regulatory proposals and legal approaches to algorithmic discrimination\footnote{In this regards, it is worth mentioning a new theory of harm that addresses the unprecedented technological capabilities of AI to cause inequality. The said \textit{artificial immutability} theory protects those algorithmic groups, treating them as \textit{de facto immutable criteria}. Immutable characters are those that we cannot change such as sex, color of skin,... \citet{wachter2022theory} suggests a wider interpretation of vulnerability under Western non-discrimination law that includes protection for algorithmic groups in the context of anti-discrimination law."}. We should note that accountability lies in the margin of discretion. When algorithmic discrimination happens, we cannot hold the systems accountable. Purely technical solutions are also not enough. The developer or deployer of the system would be held responsible in case of discrimination. Therefore, laws and policies should draw clear guidelines and offer individual protection for algorithmic discrimination. Allocation of liability is still under discussion in the European policy and law-making under the upcoming Artificial Intelligence Act (AIA), which is the first set of regulation for AI systems \cite{AIA}.  
    ML scholars explored various ways in which bias can arise and offered several bias mitigation techniques, as can be seen in \citet{mehrabi2021survey}. They have also introduced several fairness definitions. A good overview of the frequently used ones for classification can be found in \citet{verma2018fairness}. However, studies that discuss legal considerations throughout the bias mitigation pipeline are less common and are usually written by experts from other disciplines. \citet{zehlike2020} conduct such work but on a continuous rather than a discrete setting, whereas we focus on the latter. Such interdisciplinary studies are crucial as concluded in \cite{abuelyounes2020} \cite{zora207228}. Moreover, \citet{abuelyounes2020} offers a framework that aligns legal mechanisms with fairness metrics without, however, an experimental setup or more than a single case-study. \citet{kamiran2013quantifying} also consider the connection between the discrimination problem in algorithmic decision-making and anti-discrimination laws, modeling the difference between illegal and explainable discrimination. We further explore this connection in the next sections.

    %
\section{Different approaches to fairness}\label{sec:fairness}
    Different contexts, cultures, jurisdictions, and even individuals may have different perceptions of fairness. The principle of fairness is not predefined and does not have a clear and unambiguous definition. Indeed, decisions pertaining to this idea typically involve making judgment calls based on a variety of factors, unique to each case \cite{helberger2020fairest}. Concepts of equality, justice, fairness, discrimination and law are an inseparable family that cannot be understood unless examined together.

    Understanding the occurrence of discrimination in AI can be seen as a challenge in comparison with that of humans' \textit{Prima facie discrimination}\footnote{Prima Facie discrimination is easily provable due to existence of evidence in a court unless contradictory evidence proves otherwise} \cite{wachter2021fairness}. Algorithmic discrimination often has characteristics such as opacity and intangibility that make its detection and proof in a court difficult \cite{siapka2018ethical}. The lack of a \textit{comparative element} in cases of algorithmic discrimination makes matters even more complicated \cite{wachter2021fairness}. For instance, people with similar educational backgrounds might be offered different job opportunities on the basis of a characteristic that is not protected by equality laws. Traditionally speaking, if discrimination happens, one can compare and realize a sort of disadvantage in this scenario. However, in an algorithmic world, if a certain group of people who share certain characteristics do not receive a particular job posting, they would not know they were at a disadvantage. Even if they did, proving discrimination -specially without access to data- is difficult if not impossible \cite{yousefi2022notions}\cite{hacker2018teaching}.

    Algorithmic decisions\footnote{Algorithmic decisions include both automated decision-making as identified by Recital 71 and Article 22 of the GDPR (i.e., Decisions based solely on automated processing) and semi-automated decisions in which algorithms support a human decision-making process. Automated discrimination in the GDPR is limited to unfair data processing. We therefore prefer to use the term algorithmic decisions over automated decisions for the scope of this work.} also have the potential to exhibit discrimination based on new grounds because discriminatory outputs generated by algorithms could be based on less clear categories or previously unconsidered attributes. Therefore, with the exception of social groups that have traditionally been marginalized, algorithmic discrimination may fall outside the personal and material scope of the relevant non-discrimination laws, while remaining ethically unfair\footnote{Not all discrimination is illegal. Some cases of discrimination can be unethical. For instance, it has been argued that discrimination against coeliacs or vegetarians are not directly in breach of law, but can lead to morally-undesirable effects on certain groups due to a protected attribute such as health condition or philosophical beliefs\cite{sapienza2022p}.}. Algorithmic decisions can thus be an impediment to the achievement of legislative equality goals, despite not directly violating any legal norm. Existing non-discriminatory laws may thus not be enough to prevent biased practices resulting from these new categories of discrimination \cite{yousefi2022notions}.

    The study of discrimination 
    serves as one of the foundations for the mathematical definitions of fairness. \citet{srivastava2019mathematical} explain that statistical concepts of fairness require a specific metric that quantifies benefit or harm to be equal across various social groups (e.g., gender or racial groups). 
   
    As fairness is a human concept that is difficult to measure and formulate in algorithms, we should not expect algorithmic decisions to be absolutely fair, as many fairness metrics exist and none can encompass all possible meanings of fairness nor be fully adequate to all contexts. 
    Scholars have noted that fairness notions can be incompatible with one another \cite{berk2021fairness}\cite{kleinberg2016inherent}. Therefore, the choice of the fairness metric is an important part of developing and testing AI decision-making models, as it can mitigate or enhance discrimination. There is a gap in research studying fairness metrics by experiments in accordance with EU non-discrimination laws to assess their efficiency in bias mitigation and their compliance with law \cite{abuelyounes2020, zora207228}.  

    In the next subsections, we discuss fairness in EU non-discrimination laws (subsection \ref{sec:fairness_law}) and our approach to fairness in computer science (subsection \ref{sec:fairness_ML}). This then leads to section \ref{sec:exp_setup} in which we describe the experiment setup.

    \subsection{Fairness in EU Non-Discrimination Laws}\label{sec:fairness_law}
    The occurrence of discrimination through algorithmic decisions is a key challenge for societies, law and policy-makers, as the tendency to rely on such decisions inevitably grows in important instances where humans are directly impacted. In this regard, the EU is continuously attempting to keep up with the fast pace of technologies in policy and law-making. In 2019, the High-Level Expert Group released the Ethics Guidelines for Trustworthy AI that refer to AI systems being lawful, ethical and robust\footnote{The EU High-Level Expert Group on Artificial Intelligence (AI HLEG) was formed in 2018 to create ethical and trustworthy guidelines for the development of AI in Europe. The group, consisting of 52 experts, published a report in 2019 that included guidelines for transparency, accountability, and privacy in AI development, aimed at promoting innovation and growth while upholding fundamental rights and ethical values.} \cite{HLEG}. Diversity, non-discrimination and fairness is one of the seven key requirements that should be met for AI systems to be deemed trustworthy under these guidelines. In order to better understand and achieve these requirements in relation to fairness and lawfulness, we should understand fairness as viewed in law. In this section, fairness and non-discrimination under EU laws will be explored.  
    
    Laws often interconnect the concepts of fairness and equality with non-discrimination. Article 14 of the European Convention on Human Rights (ECHR) sets the right to non-discrimination "[...] on any ground such as sex, race, colour, language, religion, political or other opinion, national or social origin, association with a national minority, property, birth or other status\footnote{European Convention on Human Rights, opened for signature 4 November 1950, 213 UNTS 221 (entered into force 3 September 1953) art 14.}." As the ECHR is considered to be part of general principles of EU law, EU institutions and the Member States must respect the rights and principles enshrined in this international treaty. Since the EU has rectified the ECHR, EU laws including laws of non-discrimination must be interpreted and applied in a way that is consistent with the standards established in the ECHR. Therefore, Article 14 of the ECHR is relevant to the EU laws of non-discrimination at a constitutional level since it provides a general principle of international human rights law that must be respected and applied in EU. The EU has incorporated this principle into its own legal framework, at primary and secondary levels. 
    
    At the primary level, the non-discrimination laws in EU are enshrined in Article 21 of the EU Charter of Fundamental Rights of the European Union (CFREU)\footnote{Charter of Fundamental Rights of the European Union, [2012] OJ C 326/391 (entered into force 1 December 2009) art 21}. Everyone is equal before the law, and discrimination on the basis of certain protected characteristics, such as race, gender, age, religion, and disability is prohibited. 

    Secondary laws such as directives and decisions are important in setting out detailed rules and procedures for specific policy areas, such as non-discrimination, that are not explicitly covered in the primary law. At the secondary level, the EU has enacted several directives and regulations aimed at promoting and protecting the principle of non-discrimination in specific areas such as employment, racial and gender equality. These directives include the Racial Equality Directive\footnote{Council Directive 2000/43/EC of 29 June 2000 implementing the principle of equal treatment between persons irrespective of racial or ethnic origin, [2000] OJ L180/22.}, Employment Equality Directive\footnote{Council Directive 2000/78/EC of 27 November 2000 establishing a general framework for equal treatment in employment and occupation, [2000] OJ L303/16.}, Gender Equality Directive\footnote{Council Directive 2006/54/EC of 5 July 2006 on the implementation of the principle of equal opportunities and equal treatment of men and women in matters of employment and occupation (recast), [2006] OJ L204/23.}. These directives are addressed to the member states and are binding as to the result to be achieved. This means that member states must implement the objectives of the directive into their national legal systems, but they have some discretion as to how they do so.

    In EU non-discrimination laws, discrimination presents itself as direct, indirect, and intersectional. 

    \textbf{Direct discrimination} occurs when a person or a group of people face a less favourable treatment in comparison with others in a similar situation because of a protected characteristic. An example of direct discrimination could be a company refusing to hire a woman because of her gender\cite{bell2002anti}. 
    
    \textbf{Indirect discrimination} occurs when an apparently neutral rule, criterion or practice puts a person with a protected characteristic at a particular disadvantage compared to others. An example can be a company that requires height over 180 cm as a criteria for hiring, which would put women at a disadvantage as their average height is lower than that\cite{ellis2012eu}.

    \textbf{Intersectional discrimination} occurs when discrimination refers to the belonging of one person to several protected groups, each a victim of widespread discrimination. This overlapping affiliation can lead to different experiences of discrimination from those who belong to either group such as being a woman of color or a person with disability who also identifies as transgender\cite{atrey2019intersectional}. 
    
    Direct and indirect discrimination based on the protected characteristics in article 21 CFREU are illegal. Intersectional discrimination has been addressed in fewer instances by the law \cite{bullock2012multiple}. In some instances, discrimination may be justified by a necessary and legitimate aim when the means of achieving that aim are proportionate \cite{wachter2021fairness}\cite{moeckli2010equality}. In other words, the courts may, in certain instances, accept that different treatment has occurred but that it is acceptable and explainable. It is important to note that direct discrimination is difficult to be justified. While cases involving indirect or intersectional discrimination are much more difficult to prove but are justifiable with a necessary and appropriate aim. 
    However, any justification for indirect discrimination must be subject to strict scrutiny by the courts, and the burden of proof is on the party claiming that the discrimination is justified. This means that any justification must be carefully examined and rigorously tested to ensure that it is not based on stereotyping, prejudice, or subjective beliefs.

    To justify differential treatment, it must be shown through a proportionality test that the rule or practice in question pursues a legitimate aim, and that the means to achieve the aim (which is the reason for differential treatment) is proportionate to and necessary for achieving that aim. It must be proven to the court that there is no other means of achieving the aim and that the less favourable treatment is the minimum possible level of harm needed to achieve the aim sought, which should itself prove to be an important enough aim to justify the treatment\cite{wachter2020bias}.
    For example, a company may require job applicants to have a certain level of education or experience, but if that requirement has a disproportionate impact on people from a particular racial or ethnic group, it may be considered indirect discrimination. In such cases, the company may be able to justify the requirement if it can show that the level of education or experience is necessary to perform the job effectively and that there are no other means of achieving that aim that would have a less discriminatory impact.

    Detecting and correcting indirect discrimination can help shed light on the organizational and structural injustices in society and advocate for social change as it deals with hidden societal rules or patterns of behaviour. However, there is a lack of case-laws addressing this concept in the European Union \cite{makkonen2007measuring} 

    Illegal and explainable discrimination are well-distinct by the law. Law does not necessarily prohibit unethical discrimination that is not considered also illegal. Unethical discrimination is different unfair treatment regardless of whether it is against the law or not. Illegal discrimination is defined according to factors such as politics and religion for different jurisdictions. Unethical discrimination is more of a morally wrong action, that can vary depending on the perception of morality in people. We consider all illegal discrimination to be unethical, but not all unethical discrimination is illegal.
    For example, in many countries, it is illegal to discriminate against someone based on different protected characteristics. Such discrimination would be both unethical and illegal. However, there may be cases of discrimination which may be unethical but not necessarily illegal, for example, if an employer does not hire someone because of their physical appearance \cite{dietz2015employment}.
    

  
    When dealing with discrimination, courts must make case-specific normative choices that reflect different surrounding aspects of each case. Fairness is deeply rooted in context and can vary in every situation.
    In fact, decision-makers like judges and politicians are trained to consider \textit{substantive fairness} and \textit{procedural fairness}\footnote{Procedural fairness is the concept applied to ensure fairness in a process. In other words, the condition under which rules are followed to ensure that decisions made during the process do not discriminate e.g. fair trial or fair data processing.\cite{malgieri2020concept}} \cite{helberger2020fairest}. 
    It is a challenge to automate the same contextual fairness approach and expect an AI to comply with non-discrimination law and act fairly. \citet{wachter2021fairness} have suggested that, if used in legal and ethical decision-making, algorithmic systems could be decision-makers only when they are able to replicate the judiciary's attitude toward "contextual equality", despite the difficulty of this task.
    
    \subsubsection{Substantive Equality for algorithmic fairness}
    Interestingly, it is argued that the fundamental purpose of non-discrimination law is not only \textbf{formal equality} that aims at preventing persistent and ongoing discrimination by treating everyone equally. The purpose of non-discrimination is also to change society, policies and practices, and to create an equal starting point for everyone in order to achieve \textbf{substantive equality}. Jurisprudence of the European Court of Justice (ECJ) demonstrates that substantive equality is the aim of non-discrimination law, and that differences between groups must be acknowledged to achieve substantive equality in practice \cite{devos}.\footnote{\citet{devos} refers to several case-laws as evidence that the ECJ seeks substantive equality as the aim of the law. For instance, Case C-167/97: ex parte Nicole Seymour-Smith and Laura Perez, 1999 E.C.R. I. 623 (1999). In this case the ECJ noted the importance of a full comparison between the advantaged and the disadvantaged groups, that is interestingly also representative of the \textit{gold standard}.}
    In this view, fairness metrics address technical biases and can play a certain role in the prevention and reduction of discrimination. However, they only solve the problem of discrimination on the surface and can't address the social biases that underlie inequality \cite{wachter2020bias}.

    AI can be viewed as a technical remedy to shed light on existing inequalities. It can help build new policy interventions that aim at correcting historical biases and inequalities in the future. This requires an approach that acknowledges that status quo is often not neutral for everyone, and certain groups usually start from different unequal points, due to the historical bias they have suffered.
    
    Neither non-discrimination law, nor fair ML practices, solve the problem of social bias today. The legacy of past and present inequalities are mirrored in data that are fed to train ML models that make algorithmic decisions \cite{roselli2019managing}.

    In an attempt to understand how fairness metrics comply with non-discrimination law purposes and emphasizing on the importance of choosing fairness metrics for algorithmic decisions, \citet{wachter2020bias} define two groups of metrics: bias preserving and bias transforming. \textbf{Bias-preserving} metrics reproduce "historic performance in the outputs of the target model with equivalent error rates for each group as reflected in the training data (or status quo)". \textbf{Bias transforming} metrics are satisfied by matching decision rates between groups and do not "accept social bias as a given or neutral starting point that should be preserved, but instead require people to make an explicit decision as to which biases the system should exhibit."  The latter is therefore preferred in pursuing substantive equality goals in non-discrimination law \cite{wachter2020bias}. 
    
    However, we cannot even guarantee that humans do not suffer from prejudices and act entirely fair towards everyone, while considering the inequalities that unprivileged people have suffered. The hope is that AI systems -if developed and deployed correctly- might actually act fairer in some instances than humans, as such systems do not intentionally discriminate. For that, we could approach algorithmic fairness through mathematical definitions that have reflected philosophical and ethical concepts of human fairness to create measures for getting as close to fair as possible.
    
    The generally accepted equality concept in Western law and philosophy mirrors the Aristotelian definition of equality: “treat like cases as like” and "different cases differently" unless there is a legitimately objective reason not to do so. Formal equality is achieved on these terms. However, substantive equality is not guaranteed with simply treating everyone equally \cite{mackinnon1991reflections}. Certain attributes (e.g., gender, race) should be treated with more active attitude that makes up for social and historical unequal realities to achieve substantive equality. Indeed, substantive equality is not achieved as in the formal equality approach by ignoring the attributes based on which social groups have been treated with discrimination in the past and treating everyone the same going forward.
    Under substantive equality, the goal would be to create fair procedures using decision-making criteria that account for historical inequalities. Inequality must be viewed as a background "fact of life" for certain groups and should not be assessed individually. Groups that are historically underprivileged should therefore be put at an advantage to guarantee fairness and compensate for the historical and social inequalities. The objective is not merely to provide a better outcome for some members of a disadvantaged group. Instead, substantive equality aims to level the playing field for all participants by defining decision-making procedures and criteria in light of historical disparities. This equality concept would be met when everyone has the same starting point \cite{wachter2020bias}. 
    Fundamental structural changes are needed to shift from the merely formal view of equality, and taking an active attitude to break up the barriers which stand in the way of ideal substantive equality \cite{ellis2012eu}. 

    \subsection{Fairness in Machine Learning}\label{sec:fairness_ML}
        Algorithmic decision-making using ML is a vast field of AI, concerned with many kinds of applications. 
        The fairness metrics and bias mitigation techniques need to be adapted to each specific subdomain and type of problem. \citet{mehrabi2021survey} give a good overview of the different subdomain and the research that has been done within each area.

        From this part on, we address fairness in the canonical problem of classification, which is prevalent in fairness and AI research. It also has widely used applications and thus high chances of affecting people's lives, making a relevant case study. Many of the examples given at the beginning of the introduction are classification applications. In this study we consider the problem of fair binary classification with one binary sensitive attribute.
    
        \emph{Classification} is concerned with predicting the class (also called label) of a new observation, using the knowledge gained from previous observations for which the class is known. Those previous observations, which potentially contains bias, correspond to the data that is used to train the predictions model, also called \emph{classifier} in this context.
        Formally, the task is to learn a classifier $C$ that associates a class label $y$ to a set of attributes (or features) $F$. The problem of fair classification considers one or more protected attribute(s) (or sensitive feature), upon which no discrimination can be made.
        
        We adopt the following notation: 
        \begin{itemize}
            \item $g\in\{\text{privileged},\text{unprivileged}\}$ : protected attribute,
            \item $X$ : set of unprotected attributes,
            \item $Y\in\{+,-\}$ : actual outcome or ground truth,
            \item $\hat{y}\in\{+,-\}$ : predicted outcome.
        \end{itemize}
        We thus have $C(X_i) = \hat{y}_i$ for the $i^{th}$ instance.
        \emph{Ground truth} refers to the correct classification outcome as reflected by the relevant dataset. \emph{Predicted outcome} refers to the prediction, the result given by the classifier. The '$+$' sign represents a positive classification outcome, such as \textit{no recidivism risk} or \emph{accepted to university}, whereas the '$-$' sign represents a negative outcome, such as \textit{recidivism risk} or \emph{low income}. 
    
        As stated earlier, many approaches have been developed to increase the ability to measure biases and build fairer models. A recent survey references 91 fairness definitions and 99 bias mitigation methods just concerning classification \cite{hort2022survey}.
        Distinct metrics and methods may correspond to diverse fairness approaches or ways to address the problem, which include focusing on different stages of the model development process. 
    
        Bias mitigation methods, in all areas of ML, can be categorized as pre-, in- and post-processing \cite{dalessandro2017conscientious}.
        \emph{Pre-processing} methods are applied to the data to remove underlying bias before the model is trained on it. \emph{In-processing} methods include bias mitigation during the training process through the learning algorithm. \emph{Post-processing} methods modify a model or its results after it has been trained. It is possible to apply more that one type of technique for the same model.
    
        Some fairness definitions and their related metrics are meant to be applied on the dataset labels $Y$ to uncover their underlying bias. Metrics can also be based on the predicted outcome $\hat{y}$ to assess the bias of the model. Such metrics can thus be used if the ground truth is unavailable when auditing for fairness, for example for ownership or privacy reasons.
        The classifier bias can also be evaluated by comparing the classifier predictions $\hat{y}$ for different demographic groups and the actual outcomes $Y$, or by considering both the probability $p$ given by the classifier for an instance to belong to a certain class $c$, with $ p = P(Y_i = c | g_i, X_i)$, and the actual outcomes $Y$. Other metrics and definitions are based on similarity between individuals or on causal reasoning to evaluate the relationships between the protected attributes and the outcome \cite{hort2022survey}. 
        Fairness metrics can also be used in bias mitigation methods to optimize the model in that regard and not only with respect to accuracy\footnote{Accuracy is the rate at which the classifier makes correct predictions, with values ranging from 0 to 1. The closer to 1, the more accurate the predictions.}.
    
        Before choosing which fairness metric to use, it is important to understand the specificity of each of them and how they relate to general concepts of fairness. In this paper, we conduct our case study on Demographic Parity and use Conditional Demographic Disparity to get more insight on the results.
        
        \begin{definition}
        \textbf{Demographic Parity} (DP) is achieved when the probability to receive a favorable outcome is the same whether someone belong the privileged or unprivileged group \cite{dwork2012fairness}. 
        $$ P(\hat{y}=+ | G=\text{unprivileged}) = P(\hat{y}=+ | G=\text{privileged})$$
        \end{definition}
      
        DP is the most commonly used fairness definition for fairness research in classification \citet{hort2022survey}. It is a statistical group fairness definition based on predicted outcome. DP can also be used to analyse the dataset before training. It uncovers both direct and indirect discrimination. \citet{wachter2020bias} categorize it as bias transforming, which means it can be used to assess the level of substantive fairness of a model. It is however not a one-fits-all solution to fair algorithmic decisions and should not be chosen blindly. Since it can compare any chosen groups, including demographic groups selected according to several protected attributes, it can also be used to evaluate intersectional discrimination.
        
        The related metrics compare the positive predictions for the two groups using a difference (Statistical Parity Difference) or a ratio (Disparate Impact). They can be used as constraint for bias mitigation, which we do with the ratio in our experiment.
        
        DP has faced criticism in literature as it can lead to reverse discrimination and cannot differentiate between illegal and justifiable discrimination \cite{kamiran2013quantifying} \cite{wachter2021fairness}. It reports differential treatment (or global discrimination \cite{kamiran2013quantifying}).
        
        
        \begin{definition}
        \textbf{Conditional Demographic Disparity} (CDD) considers a model to be fair when the proportion of unprivileged people amongst those with unfavorable outcome is equal to their proportion amongst those receiving favorable outcome, considering an explanatory attributes $R$.
        $$ P(G=\text{unprivileged}|\hat{y}=+,R=r) = P(G=\text{unprivileged}|\hat{y}=-,R=r) $$
        \end{definition}
        \noindent CDD thus extends DP by adding one or more explanatory attributes $R$ \cite{kamiran2013quantifying} (also known as condition \cite{wachter2021fairness}). It has taken inspiration from conditional demographic parity \cite{kamiran2013quantifying} and is introduced by legal scholars \cite{wachter2021fairness}. The goal of this definition is to remedy some of the limitations of DP. As this choice can be made independently for each case, it also includes some component of the contextual approach to fairness that is found in the law. Additionally, it allows to discern between the part of the differential treatment that results from a justified reason (represented by the explanatory attributes) from the part that is due to illegal or unethical discrimination. The aim of CDD is to report the latter.
        
        These explanatory attributes need to be chosen by legal and/or domain experts \cite{kamiran2013quantifying}. According to \citet{wachter2021fairness}, the choice of explanatory arguments should be made by courts in each case, in accordance with the contextual approach to equality and the gold standard that is followed by the ECJ and by national courts in some case-laws. In this work, the choice of $R$ is made arbitrarily in that regard as we do not have access to such experts.
        
        Similar to DP, CDD is a statistical group fairness definition that is categorized as bias transforming \cite{wachter2020bias} and based on predicted outcome. It can also be used to evaluate direct, indirect and intersectional discrimination. We compute it as a metric taking the ratio of the two probabilities.
        
        
        
    
    
\section{Experimental Setup}\label{sec:exp_setup}

    In order to analyze the relevance of DP with regard to EU non-discrimination law while taking computer science reality into account, we analyze fairness and discrimination in real world datasets and in ML models trained on them. We consider the problem of binary classification with a binary protected attribute. In order to make relevant comparisons, we train several classification models with an in-processing bias mitigation method that impose a fairness constraint on the model. Experiments are available here\footnote{https://github.com/anonynousEAAMOsubmission2023/AIFairness\_EUlaw}. 

    \subsection{Metrics}
    
        We consider DP both as a measure to evaluate discrimination (in the ground truth and the model) and as fairness constraint that can be used in a bias mitigation training algorithm.
        
        To do so, we first look at the underlying discrimination found in datasets. Then, we train several classifiers with DP as fairness constraint and evaluate the discrimination present in their predictions. The discrimination is measured using DP and CDD. They are both computed as ratios, with a value of 1 indicating perfect fairness and 0 indicating complete discrimination. To keep this ratio between 0 and 1, we always take the lower proportion value over the higher proportion value.
        
        Since CDD should be computed for several groups (each value of attribute $R$), this leads to several values for a same model. We use the weighted average by~\citet{freedan2007} to create a summary statistic which we report in the results.  The choice of $R$ in all case studies is arbitrary for the sake of this experiment as we do not have access to experts in the appropriate fields. 
        Nevertheless, when choosing $R$, we consider the correlation of $R$ with the protected feature \textit{and} the labels as suggested by \citet{kamiran2013quantifying}. The reason is that a higher correlation between both of them could be evidence of explainable discrimination~\cite{kamiran2013quantifying}.
    
    \subsection{Datasets}\label{sec:datasets}
        We experiment with three real-world datasets that are commonly used to evaluate bias mitigation methods \cite{hort2022survey} in order to facilitate replication and comparison with other studies. In all cases, we consider \emph{gender} and \emph{race} as protected attributes, according to the social realities captured in the data and following existing literature \cite{le_quy2022datasets}.

        \subsubsection{COMPAS} The COMPAS dataset contains information about criminal offenders along with a risk score. We use the same dataset version as in the research by \citet{dressel2018compas}. Target label is \textit{two\_year\_recid}, which represents the recidivism labels (binary) collected by ProPublica Inc. and corresponds to whether the defendant has been rearrested within two years \cite{le_quy2022datasets}. This dataset is henceforth referred to as \textit{Scenario 1}. Gender is coded as binary, with women being privileged. We made the values for race binary by keeping the records of only \emph{caucasian} (privileged) and \emph{black} (unprivileged) offenders, in order to better reproduce \citet{celis2019classification}'s experiments and for ease of interpretation. After this preprocessing, the dataset contains 7 214 instances.
        
       After preprocessing, we note that 55\% of the individuals considered were labelled as not having recidivated, whereas 45\% were. This means that, when training a classifier to predict recidivism on a new unobserved set of people (test set), its prediction accuracy should be higher than 55\%. Otherwise, the decision-making process cannot be automated by the used algorithm. In other words, in ML-terminology, 55\% is the accuracy baseline.
        
        CDD is computed with attribute $R$ to be \emph{priors}, which is the number of prior criminal convictions. Since \emph{priors} is a continuous variable, discretization was necessary to compute CDD and train the classifier. We followed the approach in AIF360~\cite{bellamy2018aif360} (three categories: 0, 1-3, and >3 priors). A Chi-square Test revealed a significant weak relationships between \textit{two\_year\_recid}, and (i) $R$ ($X^{2}(2,N=7214)=572.92,\ p=0.00,\ V=.09,$), (ii) \emph{race} ($X^{2}(2,N=7214)=326.67,\ p=0.00,\ V=.07,$), and (iii) \emph{gender} ($X^{2}(10,N=7214)=112.21,\ p=0.00,\ V=.04,$).

        \subsubsection{Adult}
        This dataset \cite{Dua2019adult}, also known as "Census Income", contains the demographic characteristics of 48 842 individuals alongside the associated target labels that represent \emph{high income} (favorable outcome) or \emph{low income} (unfavorable outcome). Adult dataset is referred to as \textit{Scenario 2}. Gender is coded as binary with men being the privileged group. Race is also coded as binary, with \emph{white} as privileged and \emph{non-white} as unprivileged. We note that income of 76\% of the people is lower than 50 000\$ per year, whereas the income of the rest 24\% is higher than that. Similar to the previous scenario, 76\% serves as the accuracy baseline.
        
        To compute CDD, the explanatory attribute $R$~\cite{wachter2021fairness} is chosen to be \textit{education}. A Chi-square Test of Independence revealed a significant moderate relationship between the target and $R$ ($X^{2}(15,N=32561)=4429.65,\ p=0.00,\ V=.11.$), and a significant weak relationships between the target and (i) \emph{race} ($X^{2}(60,N=32561)=730.67,\ p=0.00,\ V=.04.$) and (ii) \emph{gender} ($X^{2}(15,N=32561)=297.725,\ p=0.00,\ V=.03.$).)

        \subsubsection{Law dataset}
        The Law School FOIA Dataset 1.1 (Law dataset) \cite{WinNT} is provided by Project SEAPHE containing admissions data of 124 557 individuals from all of the public law schools in the United States. Target variable is \emph{admit}, the labels of which are either \emph{$0$} (not admitted, i.e. the unfavorable outcome) or \emph{$1$} (admitted, i.e. the favorable outcome). The Law dataset is referred to as \textit{Scenario 3}. Gender is coded as binary with men being the privileged group. Race is also coded as binary, with \emph{white} as privileged and \emph{non-white} as unprivileged. We note that 74\% of the people where not admitted, whereas the rest did, which will serve as the baseline accuracy that an algorithm should attempt to beat.

        To compute CDD, the explanatory attribute $R$~\cite{wachter2021fairness} is chosen to be \textit{GPA}. Since \emph{GPA} is a continuous variable, discretization was necessary to compute CDD. We therefore created three categories: $<2.5$ (low score), from 2.5 to 3.4 (moderate score), and $> 3.5$ (high score). A Chi-square Test of Independence revealed significant weak relationships between the target and (i) $R$ ($X^{2}(15,N=96584)=4402.89,\ p=0.00,\ V=.08.$), (ii) \emph{race} ($X^{2}(2,N=96584)=2821.76,\ p=0.00,\ V=.06.$) and (iii) \emph{gender} ($X^{2}(2,N=96584)=636.0,\ p=0.00,\ V=.03.$).
        
        
    \subsection{Training fair models}\label{sec:training}

    We train classification models with an in-processing bias mitigation method. We use the training meta-algorithm presented in \cite{celis2019classification}, which produces an approximately fair solution, given a choice of fairness constraint and optimizing on accuracy. The fairness constraint is imposed through a function measuring the fairness level of the model and which value can't be bellow a certain threshold value $\tau$. This function corresponds to a fairness metric that is given as input. The fairness parameter $\tau$, which gives the minimal value allowed for the measure of fairness, is also given as input. Its value can range from 0 (no fairness constraint, thus no bias mitigation) to 1 ("perfect" fairness). The implementation we work with (available through AIF360 \cite{bellamy2018aif360}) uses gradient descent and considers fairness metrics using the ratio between the privileged and unprivileged groups.

    The fairness metrics according to which the model is constrained is given as input and can be one out of a large number of metrics, larger than for other comparable algorithm. This allows for a relevant comparison between the effect of several different metrics on the resulting models, without being impacted by other differences in the algorithm training.


    There are several reasons that motivated us to choose this bias mitigation meta-algorithm. First, the authors provide mathematical proofs that guarantee that the fairness constraints are satisfied (up to a certain controlled error) while the classification model is optimized. Second, their approach handles linear-fractional constraints rather than strictly linear which is mostly done in the literature. This means that fractional metrics (i.e. non-linear) such as the False Discovery Rate can be considered, consequently increasing the number of available fairness metrics. Third, they have written this constrained classification problem as a linear program, allowing them to develop a computationally efficient algorithm (of polynomial complexity).
    
    We use this meta-algorithm to create several models for each scenario, using DP as fairness constraint, considering successively gender and race as sensitive attributes. For each scenario and each sensitive attribute, we compute models for input $\tau$ ranging from 0 (no fairness constraint) to 1 (constraint to aim at perfect fairness). We refer to each of the models created as \emph{metaclassifier}, following the name of the learning algorithm. We use 10-fold cross-validation\footnote{With this method, we repeat the training process 10 times using different samples from the same dataset to derive more representative results.} and report the average of each computed metric. We split both datasets into training ($70\%$) and test set ($30\%$), selected uniformly at random using the method in AIF360\cite{bellamy2018aif360} which results in a stratified split. We use the open-source implementation of the meta-algorithm provided by AIF360 \cite{bellamy2018aif360} to ease reproduction, with our own pre-processing.

\section{Results and Discussion}\label{sec:results}
    In this section, we present the experimental results and interpret them considering both the computer scientists' and legal scholars' perspectives.

    \subsection{Choosing a $\tau$ value}\label{sec:results_tau_choice}

\begin{figure}
    \centering
        
        \begin{subfigure}[b]{0.8\linewidth}
            \centering
            \includegraphics[width=1.00\textwidth]{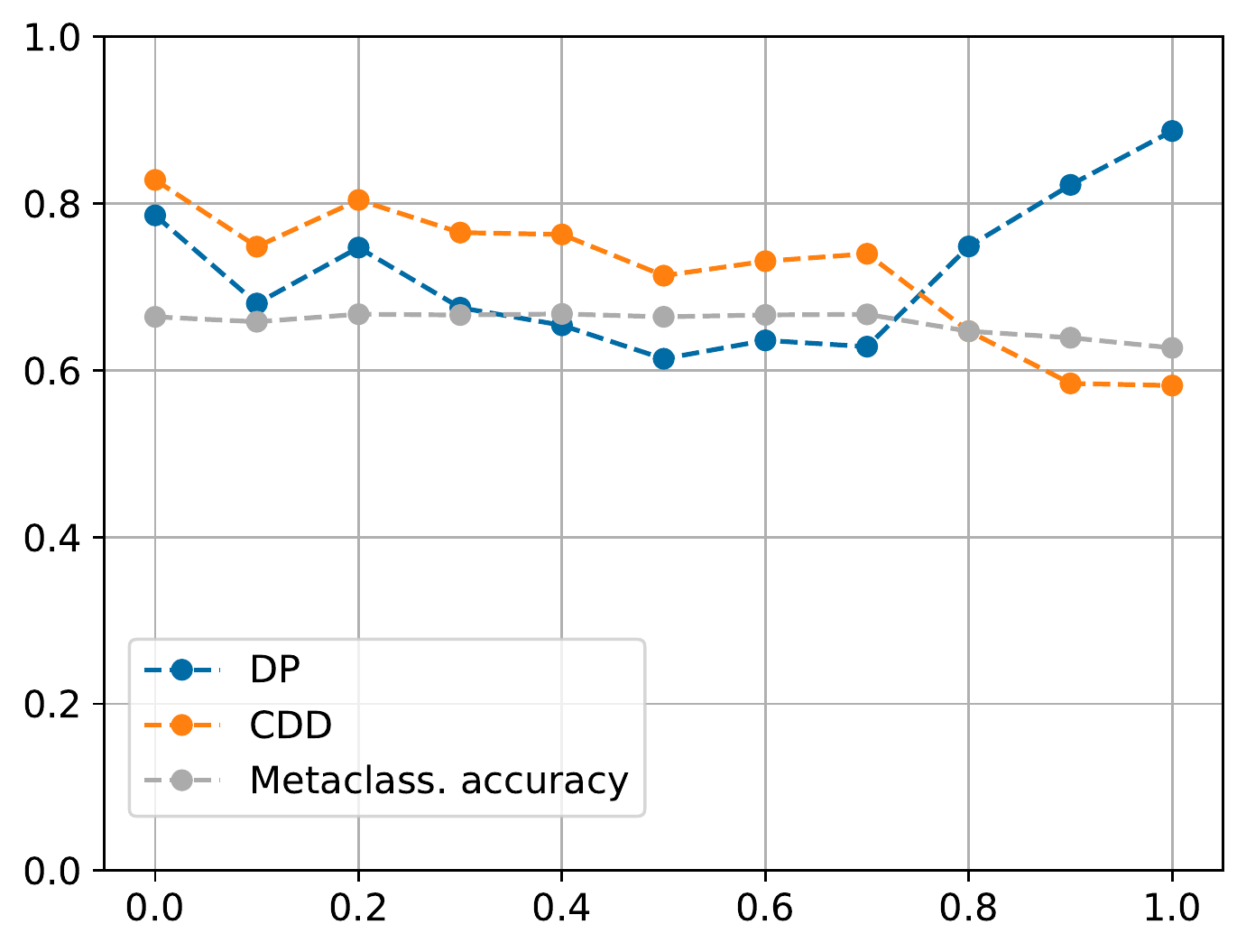}
            \caption{Scenario 1: COMPAS dataset \& gender}
            \label{fig:compas_twoyearsgender}
        \end{subfigure}
        \begin{subfigure}[b]{0.8\linewidth}
            \centering
            \includegraphics[width=1.00\textwidth]{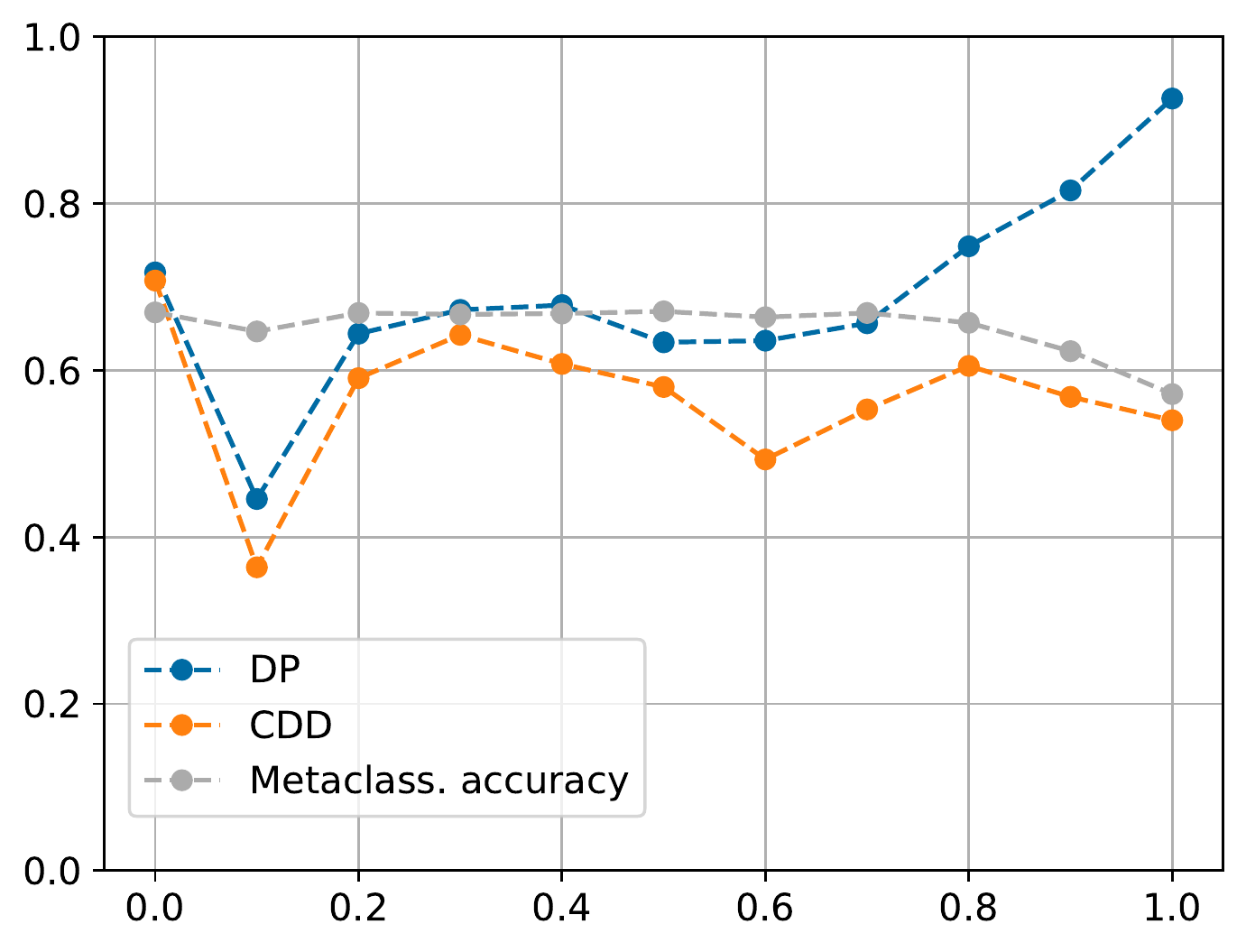}
            \caption{Scenario 1: COMPAS dataset \& race}
            \label{fig:compas_twoyearsrace}
        \end{subfigure}
        \caption{Fairness metrics and accuracy results for Compas dataset after 10-fold cross-validation.}
    \label{fig:compas}
    \end{figure}

\begin{figure}
    \centering
        \begin{subfigure}[b]{0.8\linewidth}
            \centering
            \includegraphics[width=1\columnwidth]{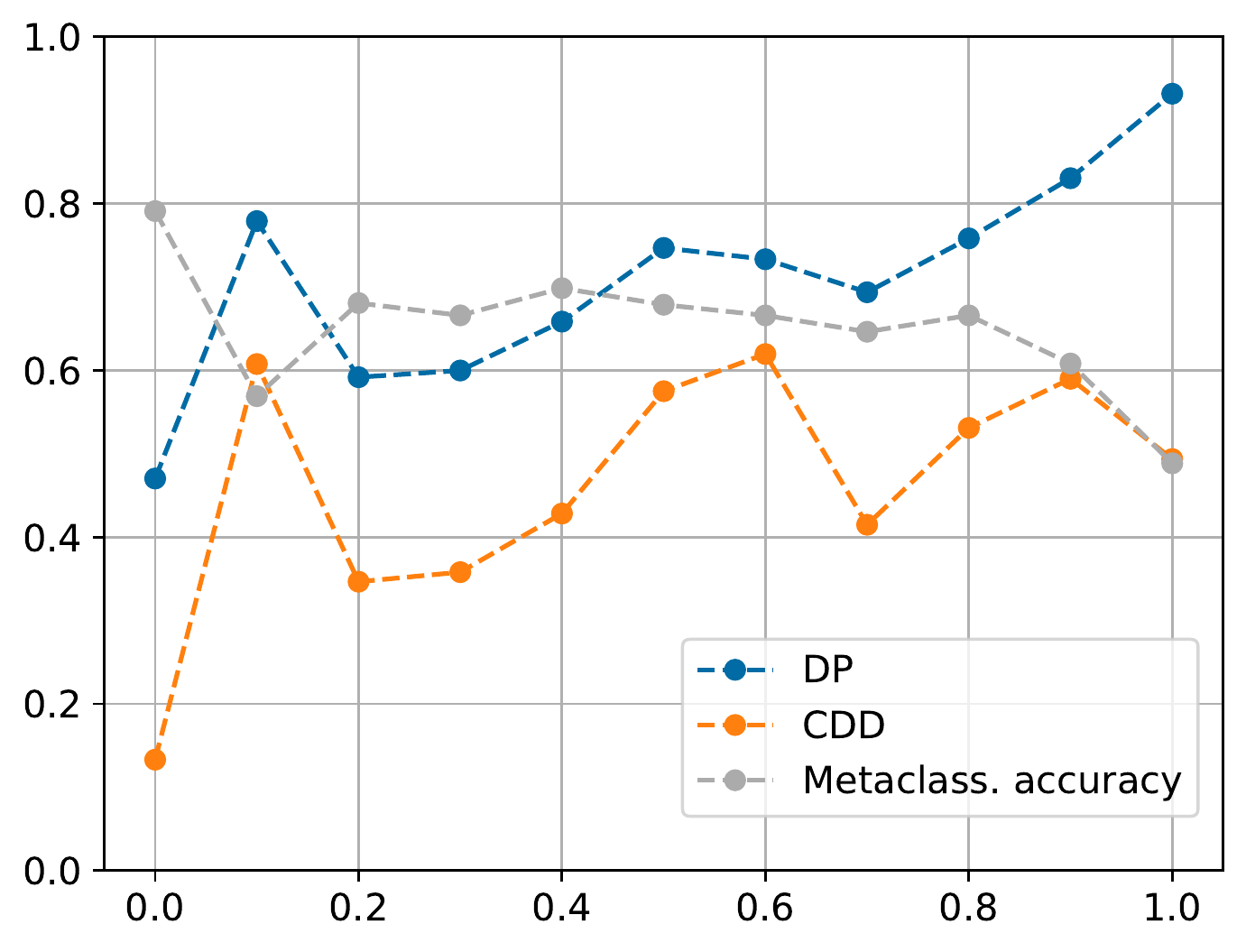}
            \caption{Scenario 2: Adult dataset \& gender}
            \label{fig:adultsrsex}
        \end{subfigure}
        \begin{subfigure}[b]{0.8\linewidth}
            \centering
            \includegraphics[width=1.00\textwidth]{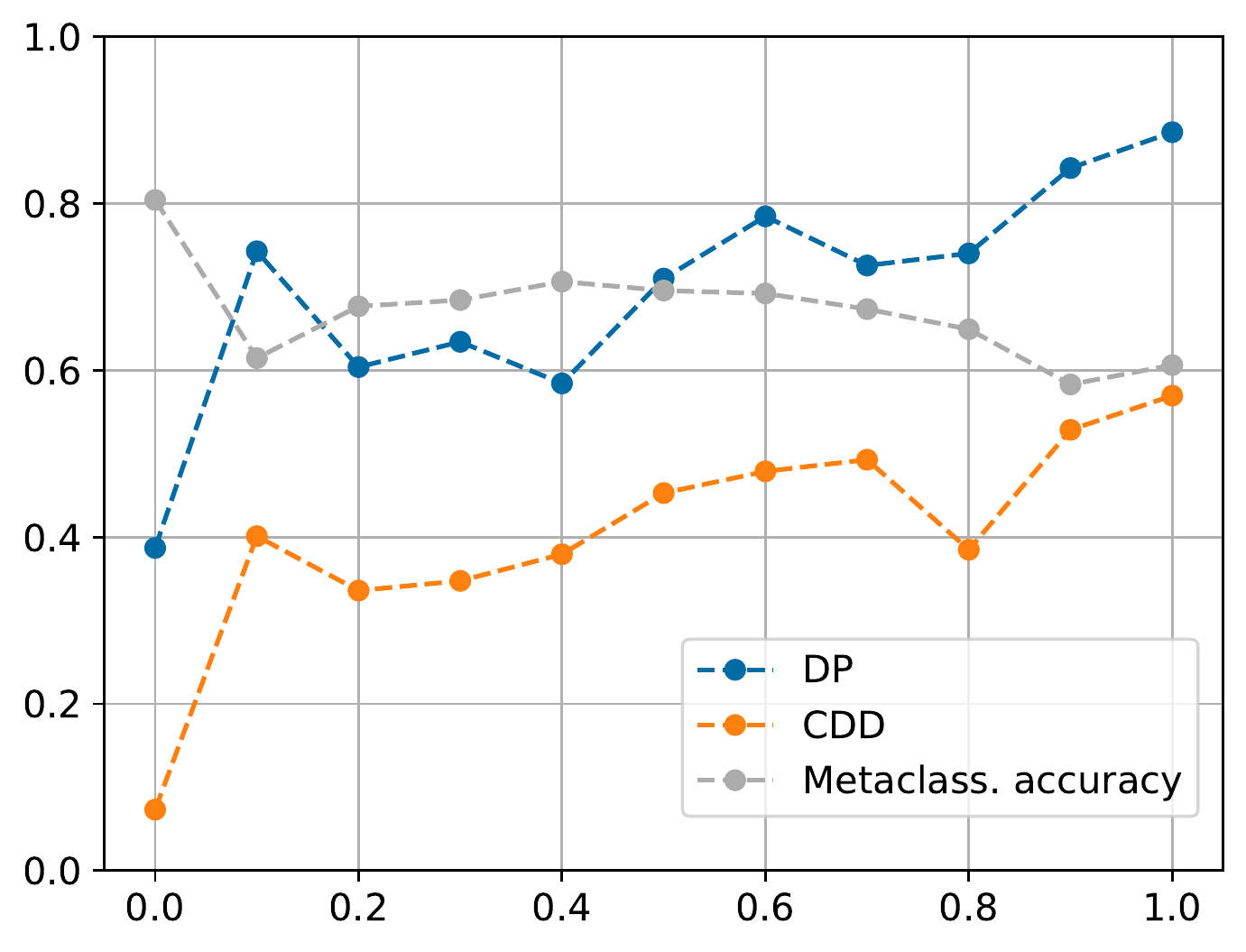}
            \caption{Scenario 2: Adult dataset \& race}
            \label{fig:adultsrrace}
        \end{subfigure}
    \caption{Fairness metrics and accuracy results for Adult dataset after 10-fold cross-validation.}
    \vspace{-1.5em}
    \label{fig:adult}
    \end{figure}

\begin{figure}
    \centering
        \begin{subfigure}[b]{0.8\linewidth}
            \centering
            \includegraphics[width=1\columnwidth]{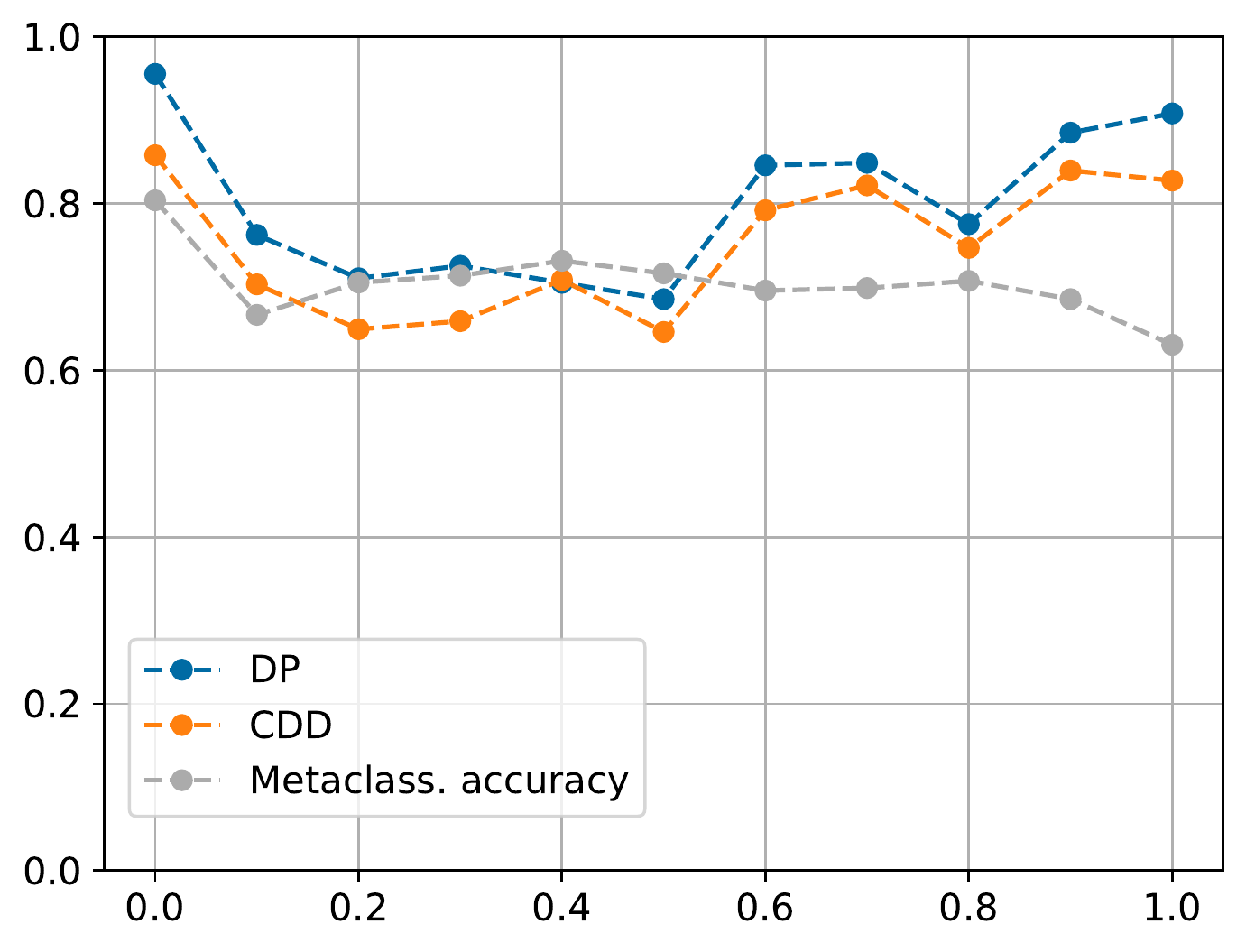}
            \caption{Scenario 3: Law dataset \& gender}
            \label{fig:lawsrsex}
        \end{subfigure}
        \begin{subfigure}[b]{0.8\linewidth}
            \centering
            \includegraphics[width=1.00\textwidth]{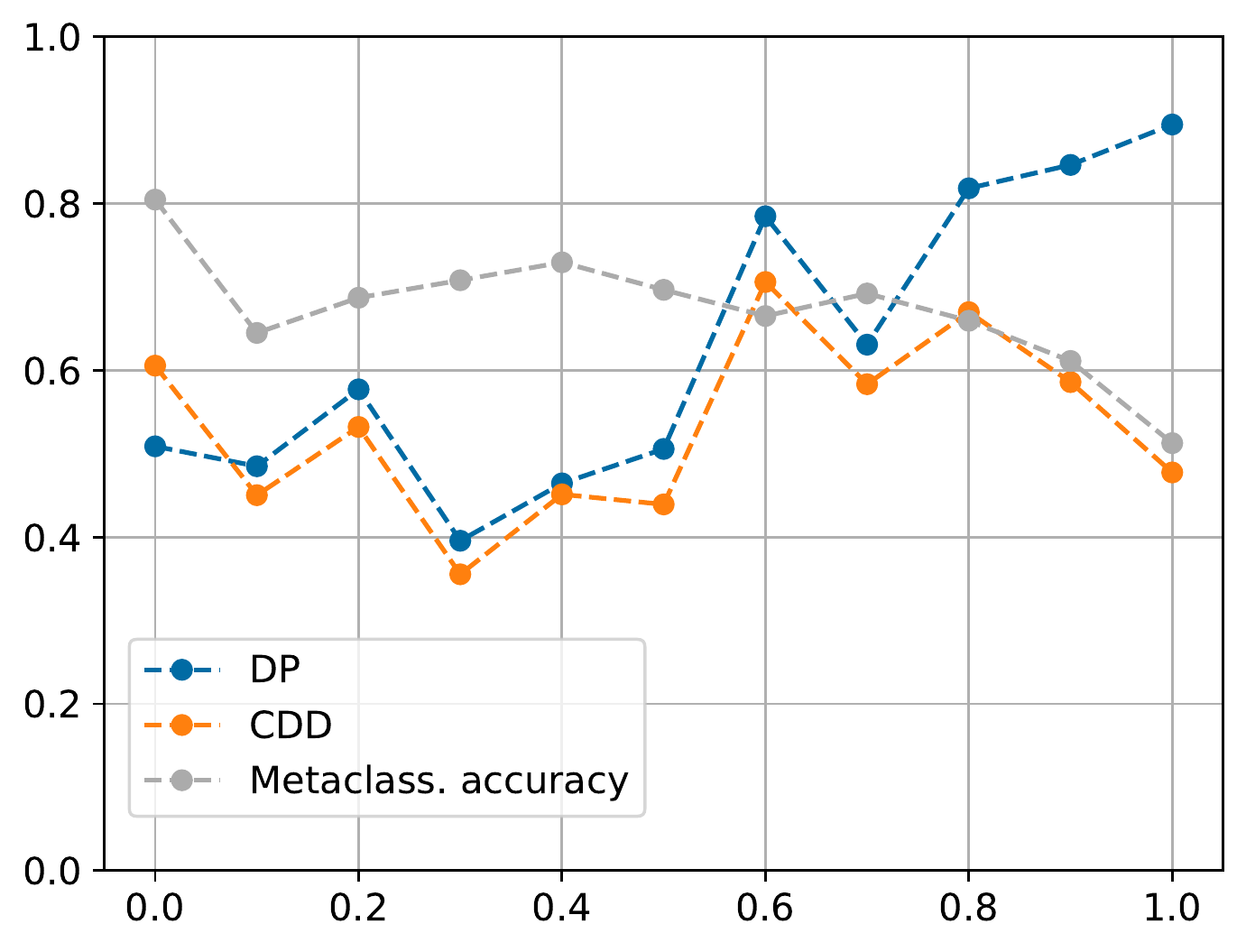}
            \caption{Scenario 3: Law dataset \& race}
            \label{fig:lawsrrace}
        \end{subfigure}
    \caption{Fairness metrics and accuracy results for Law dataset after 10-fold cross-validation.}
    \vspace{-1.5em}
    \label{fig:law}
    \end{figure}

    Figures \ref{fig:compas}, \ref{fig:adult} and \ref{fig:law} present the fairness and accuracy values obtained for the predictions of the metaclassifiers with different levels of constraint to achieve fairness (i.e. different values of $\tau$). However, most real-life scenarios require the choice of a unique classifier. We thus considered specific values of $\tau$ for this experiment, selected in accordance with the contextual approach to equality in non-discrimination law and considering both the accuracy and fairness values.
    
    A way to approach this would be to set a fixed threshold, such as the 80\% rule set in the 1978 Uniform Guidelines on Employee Selection Procedures\footnote{This guideline states "a selection rate for any race, sex, or ethnic group which is less than four-fifths (4/5) (or eighty percent) of the rate for the group with the highest rate will generally be regarded by the Federal enforcement agencies as evidence of adverse impact, while a greater than four-fifths rate will generally not be regarded by Federal enforcement agencies as evidence of adverse impact." UNIFORM GUIDELINES ON EMPLOYEE SELECTION PROCEDURES (1978), 29 C.F.R. §1607.4 (2018)} \cite{feldman2015certifying}. However, the U.S Supreme Court has shown resistance in adapting a "rigid mathematical formula" to define disparate impact \cite{1988watson}. The European Commission's proposed AIA introduces impact assessments accompanied by codes of conduct and externally audited compliance. This regulation does not yet mention any obligatory margin of threshold for the trade-off between fairness and accuracy. The classification of metrics into bias transforming and preserving given by \citet{wachter2020bias} could be a starting point to be introduced in regulation, together with a proportionality test and the need for justification of the metric. 
    
    We said in section \ref{sec:fairness_law} that substantive equality includes considering that status quo is often not neutral, as certain groups usually start from different unequal points, due to the historical bias they have suffered. For that, choosing a specific threshold would indeed be difficult with a rigid mathematical formula. However, the existence of such threshold would enable legal operators to promote and discuss the outputs of algorithmic decisions. Therefore, it would be ideal to approach it legally with a margin to respect. The margin of this threshold could be introduced in the AIA to prevent illegal discrimination in algorithmic decisions. In that case, developers could follow the legal guidelines and if discrimination happens, in case the developers' compliance is presumed, the burden of proof would be reversed. In other words, if regulations introduce a margin of threshold that should be respected, and if developers stay within that margin, illegal discrimination could idealistically be prevented. In the case of its occurrence, it would be the duty of the plaintiff to prove discrimination has happened. Conversely, if the legal guidelines are not respected, discrimination would be presumed to have occurred.
    
    This inclusion of margins in legal texts would thus be beneficial for AI developers and deployers, as they would know what is expected of them to avoid legal repercussions. It would also be advantageous for the citizens who are subject to AI systems. First because it could help reduce the occurrence of discrimination in algorithmic decisions, thus protecting them from harm. Second because it could make it easier for individuals and civil society to bring a case of discrimination to court, leading to compensation and correction of harm if judged appropriate.
    
    Taking the said example from U.S. legal system into account while acknowledging the differences between the U.S. and EU law, we selected a relevant threshold for each scenario individually to approach the contextual requirement of EU law. For the sake of the experiment, the choice was made arbitrarily with the goal to obtain a good trade-off between accuracy and fairness. In practice, the optimal $\tau$ should be chosen by legal practitioners, developers and ethicists who are familiar with the classification problem at hand.
    
    \subsection{Evolution of fairness and accuracy with $\tau$}\label{sec:results_tau_evol}
    
    We can see in Figures~\ref{fig:compas}, \ref{fig:adult} and \ref{fig:law} the fluctuations of accuracy and fairness in the predictions of models created with different minimal fairness values $\tau$ ranging from 0 to 1. Following, we analyze the results and describe how we choose the most suitable $\tau$ value per scenario.

    \subsubsection{Scenario 1} 
    Figure~\ref{fig:compas_twoyearsgender} gives the results for the models trained on COMPAS with gender as sensitive attribute. The meta-classifier fairness constraint is thus DP with the underprivileged group being men and the privileged group being women.
    
    At $\tau=0$, DP is 0.79 with a 66\% accuracy.
    It is worth noting that accuracy remains relatively stable as the fairness constraint $\tau$ gradually increases. We can consider that the fairness-accuracy trade-off is always satisfactory in this scenario, which simplifies the process of choosing a proper constraint value, at least in that regard.

    As explained in section \ref{sec:fairness_ML}, the fairness value given by DP considers the differential treatment, or global discrimination. CDD, on the other hand, gives a fairness value that reflects the discrimination that is considered unethical, under the assumption that the explanatory attribute used in CDD is well chosen and sufficient. The difference between DP and CDD values is related to the level of justified or explainable discrimination due to the effect of the explanatory attribute \textit{priors} on the result. 
    We can see in Figure~\ref{fig:compas_twoyearsgender} that CDD follows a similar trend as DP until $\tau=0.7$, then decreases as $\tau$ rises to 1. Since CDD is slightly higher than DP up to $\tau=0.7$, but still relatively similar to it, this means that most of the discrimination captured by DP is unethical with only a small part being explainable, considering the assumption that the explanatory attribute used in CDD is well chosen and sufficient. However, past $\tau=0.7$, CDD decreases while strongly DP increases, which is caused by the difference in the distribution over the explainable attribute \emph{priors}. This difference indicates that DP is forcing equal treatment when there would be a justifiable reason not to do so, thus possibly resulting in undesired reverse discrimination. This example illustrates how imposing too much constraint with regard to one fairness definition can have a detrimental effect according to another definition of fairness.

    Since the fairness of the model's prediction is already relatively high when it is trained without bias mitigation, based on both DP and CDD, the mitigation is only effective when the constraint value $\tau$ is greater than 0.8. When $\tau=0.9$, DP is 0.81 with a 64\% accuracy, which is only a slight increase compared to $\tau=0$.
    This is also when reverse discrimination could start happening. Thus, bias mitigation based on DP for this model might not be necessary in real-life. Deciding whether it is legally and ethically acceptable that the rate of men with a low recidivism risk is 80\% of the rate of women with low recidivism risk should be left to experts familiar with the context of this scenario. Such a decision should also take into account the potential impact of reverse discrimination and whether such an effect would be unacceptable, tolerable, or even desirable in the specific context.

    In the same scenario, when \emph{race} is chosen as protected attribute (see Figure \ref{fig:compas_twoyearsrace}), fairness is slightly decreased as compared to \emph{gender} for both DP and CDD at about 0.72 at $\tau=0$. Accuracy has a stable trend slightly decreasing from 66\% to 58\% as $\tau$ increases. This time, the trend of DP and CDD is reversed, with CDD being slightly lower than DP. DP reflects most unethical discrimination, but not the full extent of it. This difference increases with the strength of the constraint on DP, leading to a correction of discrimination that becomes less fair according to CDD when $\tau > 0.8$.
    
    After $\tau=0.6$, the trends DP and CDD follow begin to grow further apart. Choosing the optimal $\tau$ value here is related to the context and the metric that might be more applicable: a metric that attempts to balance the ratio between favored white and black people (DP), or a metric that attempts to balance the ratio between favored black people with less prior crimes and black people with more prior crimes? In this scenario, we arbitrarily choose the former, therefore, optimal $\tau$ is considered to be 0.9.

    \subsubsection{Scenario 2} The results concerning the Adult dataset are presented in Figure \ref{fig:adult}. We observe slightly higher values of fairness with regard to gender bias compared to racial bias, but overall similar trends for both considered metrics. The level of fairness with regard to both protected attributes is much lower than in \textit{Scenario 1}. We can thus expect that the bias mitigation process would be more useful, which is confirmed by the results.
    
    When $\tau=0$, DP reports a fairness value of 0.48 towards \emph{gender} and 0.39 towards \emph{race}, while CDD reports 0.15 for fairness towards \emph{gender} and 0.10 towards \emph{race}. The gap between those values can be interpreted as the part of differential treatment explained by the explanatory attribute \textit{education}. Since this gap remains relatively constant for every value of $\tau$, this indicates that the parts of global discrimination due to respectively unethical and explainable discrimination stay similar with different levels of constraint. However, this changes for the model aiming at perfect DP fairness regarding gender ($\tau = 1$ in Figure \ref{fig:adultsrsex}), where the difference increases. This indicates either reverse discrimination or an otherwise inadequate correction of the global discrimination that doesn't correctly match the unethical discrimination.
    
    Classification accuracy starts at 80\% with no fairness constraint and fluctuates between 60-70\% as the fairness constraint is gradually increasing. This is considered a weak performance since the majority label (\textit{low income}) is 76\% of the sample. In other words, the algorithmic decision-making process is not efficient given this dataset and classifier. The non-monotonic characteristic of the results is in part due the 10-fold cross-validation and in part to the training process that uses a random sample drawn out of the distribution of the input data and not the whole dataset.

    For this scenario, we consider the models with $\tau = 0.6$ for gender and  $\tau = 0.7$ for race to be the most appropriate, considering the trade-off between accuracy and fairness. This case-study offers a fine example of the fairness constraint gradually removing the bias expressed by DP as $\tau$ increases, with a similar improvement of CDD and an overall decrease of accuracy. The result for DP and accuracy align well with those reported by ~\cite{celis2019classification}.

    \subsubsection{Scenario 3} The results concerning Law dataset are presented in Figure \ref{fig:law}. When $\tau = 0$, accuracy is 80\% and DP is almost 1 for the case of \emph{gender}, and about 0.51 for \emph{race}. When it comes to \emph{gender} the decision-making process is already fair, therefore the bias mitigation method seems to only decrease fairness. This makes sense since the fairness constraint is considered with a minimal required value for DP, which is approximately respected by the produced models. Also, the training algorithm used is not exactly the same when bias mitigation is applied or not, which explains part of the difference in the results.
    CDD of \emph{gender} is also relatively high, at 0.86. CDD is slightly lower that DP and it follows the exact same trend as DP. This means that almost all the global discrimination is explainable following the GPA of the law students. Finally, Moving on to \emph{race}, it is evident that a fairness constraint should be added in the decision making process. We can clearly see the global increase of DP with stronger fairness constraints.

    Considering those results, we consider that an optimal constraint in this scenario would be $\tau = 0.6$. Indeed, at that level, both DP and CDD are much higher than when no fairness constraint is used. They also have quite similar values ($0.79$ and $0.7$ respectively), which indicates that almost all of the discrimination removed is unethical and that there can be almost no reverse discrimination. Accuracy is still high enough and decreases for higher values of $\tau$. When $\tau > 0.7$, we can also observe that DP and CDD start following different trends, with the former gradually increasing and the latter decreasing. This leads to the same concern as in \textit{Scenario 1} of a too strong constraint on DP potentially leading to reverse discrimination. However, depending on the context and application of the classifier, such reverse discrimination could be a desirable effect in order to compensate for the existing inequalities. The choice of the proper model and bias mitigation parameters is thus again highly context-dependent and should take many factors into account.


    \subsection{Effectiveness of bias mitigation}\label{sec:results_metrics}

    Table~\ref{tab:compare} presents the metric results for DP and CDD in three different situations. The first is pre-training where we compute DP and CDD on the test set using the actual target labels. This serves as our ground truth. The second is DP and CDD computed on the prediction of the metaclassifier with no bias mitigation ($\tau=0$). It thus behaves as a conventional gradient-based classifier. Finally, DP and CDD computed on the predictions of the metaclassifier with the selected value of $\tau$ where bias metrics generally perform better than $\tau=0$ while accuracy is still at acceptable levels. This corresponds to the models we have selected in the previous section. The following observations can be made based on these results.

    \begin{table}[!ht]
    \caption{Fairness results per dataset on the test set (ground truth) and predictions (referred to as Pred.).}
    \label{tab:compare}
    \resizebox{1.0\linewidth}{!}{
    \begin{tabular}{|c|c|c|c|c|c|c|c|c|c|}
        \hline
        \multicolumn{1}{|c|}{} & \multicolumn{3}{c|}{Compas: Scenario 1} & \multicolumn{3}{c|}{Adult: Scenario 2} & \multicolumn{3}{c|}{Law: Scenario 3} \\
        \cline{2-10}
        \multicolumn{1}{|c|}{} & \begin{tabular}{@{}c@{}}Test set \\ {} \end{tabular} & \begin{tabular}{@{}c@{}}Pred. \\ $\tau=0$\end{tabular} & \begin{tabular}{@{}c@{}}Pred. \\ $\tau=0.9$\end{tabular} &  \begin{tabular}{@{}c@{}}Test set  \\ {} \end{tabular} & \begin{tabular}{@{}c@{}}Pred. \\ $\tau=0$\end{tabular}  & \begin{tabular}{@{}c@{}}Pred. \\ $\tau=$best\textsuperscript{2}\end{tabular} &  \begin{tabular}{@{}c@{}}Test set  \\ {} \end{tabular} & \begin{tabular}{@{}c@{}}Pred. \\ $\tau=0$\end{tabular}  & \begin{tabular}{@{}c@{}}Pred. \\ $\tau=$best\textsuperscript{3}\end{tabular}\\
        \cline{2-10}
        \multicolumn{1}{|c|}{SA\textsuperscript{1}} & \multicolumn{9}{c|}{Demographic Parity (DP)}   \\
        \hline
        gender & \cellcolor{blue!15} 0.93 & \cellcolor{blue!15} 0.79 & \cellcolor{blue!15} 0.82 & \cellcolor{green!15} 0.35 & \cellcolor{green!15} 0.47 & \cellcolor{green!15} 0.73 & \cellcolor{blue!15} 0.98 & \cellcolor{blue!15} 0.96 & \cellcolor{blue!15} 0.85\\
        
        race & \cellcolor{green!15} 0.89 & \cellcolor{green!15} 0.72 & \cellcolor{green!15} 0.82 &  \cellcolor{green!15} 0.6 & \cellcolor{green!15} 0.39 & \cellcolor{green!15} 0.73 & \cellcolor{green!15} 0.76 & \cellcolor{green!15} 0.51 & \cellcolor{green!15} 0.79\\
        \hline

        \cline{2-10}
        \multicolumn{1}{|c|}{} & \multicolumn{9}{c|}{Conditional Demographic Disparity (CDD)}       \\
        \hline
        gender & \cellcolor{red!15} 0.8 & \cellcolor{red!15} 0.83 & \cellcolor{red!15} 0.58 & \cellcolor{green!15} 0.35 & \cellcolor{green!15} 0.13 & \cellcolor{green!15} 0.62 & \cellcolor{blue!15} 0.95 & \cellcolor{blue!15} 0.86 & \cellcolor{blue!15} 0.82\\
        
        race & \cellcolor{red!15} 0.8 & \cellcolor{red!15} 0.71 & \cellcolor{red!15} 0.57 & \cellcolor{green!15} 0.61 & \cellcolor{green!15} 0.07 & \cellcolor{green!15} 0.49 & \cellcolor{green!15} 0.85 & \cellcolor{green!15} 0.61 & \cellcolor{green!15} 0.71\\
        \hline
    \end{tabular}
    }
    \scriptsize\textsuperscript{1} Sensitive attribute.
    \scriptsize\textsuperscript{2} Here, $\tau$ is $0.6$ for gender, $0.7$ for race.
    \scriptsize\textsuperscript{3} $\tau$ is $0.7$ for gender, $0.6$ for race.
    \scriptsize\textsuperscript{4} \colorbox{blue!15}{Blue} cells refer to cases where fairness was high to begin with and while classifying with or without fairness constraint, fairness did not drop more than 0.15. \colorbox{green!15}{Green} cells refer to cases where fairness mitigation either maintained or improved fairness as compared to the test set. \colorbox{red!15}{Red} cells refer to cases where bias mitigation reduced fairness by more than 0.2.
    
    \end{table}
    
    In \textit{Scenario 1}, the level of bias against gender as computed using the test set (ground truth) is 0.93. When the metaclassifier is trained without any fairness constraint ($\tau = 0$), then fairness decreases to 0.79. This is a good example of how technical bias can arise in a model even when the training data is fair. When the fairness constraint is set to 0.9, then fairness slightly raises by 0.03 based on DP, while roughly preserving its classification power. However, CDD plunges to 0.58, which is evidence of unethical discrimination in the decision-making system. With regard to the protected attribute \textit{race}, fairness in the test set is 0.89 based on DP which could also be considered relatively high. Next, the classifier reduces fairness to 0.72 when making predictions after being trained without bias mitigation, whih means that again technical bias arose in this model. Consequently, when $\tau$ is set to 0.9, fairness increases to 0.82, which could be a fine example of a successful bias mitigation process. However, CDD is gradually decreasing from 0.80 in the ground truth, to 0.71 when predicting without a fairness constraint, and finally to 0.57 when $\tau=0.9$. Technically, the difference between DP and CDD at $\tau=0.9$ is caused by the fact that the distribution across races (measured by DP) is different from the distribution within the unprivileged race (measured by CDD). Practically, it is up to the decision-makers to decide which measure they wish to use, because they cannot satisfy both at the same time based on the data. If the decision-makers wish to emphasize on the students' GPA scores ($R$) which might nevertheless reflect historical and systemic inequalities, then CDD should be chosen. On the other hand, if the decision-makers wish to grant equal access to the underprivileged race regardless of their GPA, then DP should be chosen toward the substantive equality goal.

    In \textit{Scenario 2}, all cases show a fairness increase by more than 0.25 from $\tau = 0$ to $\tau = best$. Three out of four cases exhibit a fairness increase from the test set to $\tau = best$. Only in the case of CDD and \emph{race}, there is a 0.12 fairness decrease. However, in the same case, there is already a substantial improvement in fairness from $\tau = 0$ to $\tau = best$ which qualifies this case as a bias mitigation success. However, in \textit{Scenario 2}, there is evidence of unethical discrimination that is not properly corrected since CDD is generally lower than DP reaching at most a moderate 0.58 amount of fairness. 
    
    Finally, bias mitigation worked well when it comes to \textit{race} and DP in \textit{Scenario 3}. Bias mitigation was also successful in \textit{race} and CDD as demonstrated by the increase of CDD when it comes to \textit{race} from $\tau = 0$ to $\tau = best$. Bias mitigation was not able to reach the levels of the ground truth in that case. Nevertheless this is an encouraging result towards the goal of restoring formal equality.


    A result that can be drawn from the above is that bias mitigation can be sufficient in some situations. First, it can be sufficient in settings where the system should be designed to replicate bias (such as diagnostic tools) to comply with formal equality. This is demonstrated by the case study of Law dataset. Second, bias mitigation considering fairness constraint with bias transforming metrics such as DP can be successful in settings where the system should create decisions fairer than the ground truth (substantive equality). The Adult dataset is a good example of this. However the metrics do not align in all cases as clearly shown by Compas dataset. Hence, the use of any metric should be objectively justified under a proportionality test \cite{wachter2020bias}. Our experiments also highlight the influence of the context and how important it is to take it into consideration for bias mitigation.
    
    We should also note that ground truth and the complexity of reality may be hard to capture in data, since it can never encompass the completeness of the real world. Our contribution lies in analyzing and discussing existing fairness approaches and legislation to allow for more informed choices and understanding on the way to address bias in ML models. We do not claim to 'fix' bias in society or in AI systems relying on techno-solutionism. Rather, we talk about mitigation, i.e. reducing bias and harm that can be caused by automated decision-support systems. It is imperative to involve human experts into the process and adopting a case-specific approach in the choice of explanatory attributes and fairness constraint. We acknowledge that ethically reflecting on the aim of ML development could be a valuable extension of our work that would, however, require further expertise from a sociological perspective.

    We acknowledge that including a single explanatory attribute ($R$) is a limitation of this study. Choosing the optimal $R$ is not the focus of our work, and the fact that we did not have access to experts regarding the datasets and scenarios only reinforces this choice. In the future, we will focus on including and discussing more bias metrics. The same rationale holds for choosing the best minimal value $\tau$ for the fairness constraint and the trade-off between accuracy and fairness. Consequently, goal of our work is to bridge the gap between developers and legal scholars by analysing different scenarios in terms of fairness using promising existing approaches.

    \section{Final Remarks}\label{sec:ccl}
    In this article, we explored existing research on the definition of the concept of fairness in law and computer science. Using legal informatics as our methodology and through three scenarios (datasets), we assessed the mutual contribution of computer science and EU non-discrimination legal framework in bias mitigation and prevention of discrimination in ML-driven driven algorithmic decisions. More specifically, we measured fairness using two metrics, Demographic Parity (DP) and Conditional Demographic Parity (CDD) in different stages of a classification process. The classifier we used is introduced by \citet{celis2019classification} whose work we replicated and extended. In our results, we compare the fairness on the test sets with the fairness on the predictions of different classifiers trained with DP as a fairness constraint. 

    Our experiments confirm the significance of the choice of metrics in bias mitigation, as they should be interpreted in the context of formal or substantive equality of the law. This significance is also valid for the choice of the notion(s) of fairness that is (are) applicable to the domain and the appropriate margin for bias mitigation thresholds, such as $\tau$. 

    Based on our numeric results, we conclude that the metaclassifier performed best in the scenario involving Adult and Law datasets. In this case, DP and CDD aligned well up to a certain extent. In alignment with previous work, we noted that in case of using CDD, the condition for discrimination is not easy to set without any input from legislative or judiciary bodies or legal experts. We therefore emphasize on the urgent need for collaboration between computer scientists and the legal community in approaching fairness in ML applications. 

    Acknowledging that our conclusions are limited by the low number of chosen case studies and fairness metrics, future work will be expanded on evaluating more metrics, using other datasets and considering several sensitive attributes. Further, the complexity of discrimination - especially that of indirect and intersectional discrimination - and new sources of
    algorithmic discrimination, require conducting a more in-depth legal informatics research in the future.    
    
    We should also note that unfair predictions made by ML systems are not necessarily the result of inaccurate or incomplete data. They reflect the biased and unequal reality of the world as it is, in which ML is deployed. Individual, social and institutional changes are required to truly resolve the problem of discrimination. Algorithmic decision-making could produce accurate classification results while preserving fairness using the appropriate metrics and with the appropriate supervision, but the social problem will still prevail if we stop at technical solutions.
    
    \bibliographystyle{ACM-Reference-Format}
    \bibliography{fairness}


\begin{thebibliography}{54}


\ifx \showCODEN    \undefined \def \showCODEN     #1{\unskip}     \fi
\ifx \showDOI      \undefined \def \showDOI       #1{#1}\fi
\ifx \showISBNx    \undefined \def \showISBNx     #1{\unskip}     \fi
\ifx \showISBNxiii \undefined \def \showISBNxiii  #1{\unskip}     \fi
\ifx \showISSN     \undefined \def \showISSN      #1{\unskip}     \fi
\ifx \showLCCN     \undefined \def \showLCCN      #1{\unskip}     \fi
\ifx \shownote     \undefined \def \shownote      #1{#1}          \fi
\ifx \showarticletitle \undefined \def \showarticletitle #1{#1}   \fi
\ifx \showURL      \undefined \def \showURL       {\relax}        \fi
\providecommand\bibfield[2]{#2}
\providecommand\bibinfo[2]{#2}
\providecommand\natexlab[1]{#1}
\providecommand\showeprint[2][]{arXiv:#2}

\bibitem[198(1988)]%
        {1988watson}
 \bibinfo{year}{1988}\natexlab{}.
\newblock \bibinfo{title}{Watson v. Fort Worth Bank \& Trust}.
\newblock , \bibinfo{numpages}{977}~pages.
\newblock


\bibitem[Abu-Elyounes(2020)]%
        {abuelyounes2020}
\bibfield{author}{\bibinfo{person}{Doaa Abu-Elyounes}.}
  \bibinfo{year}{2020}\natexlab{}.
\newblock \showarticletitle{Contextual Fairness: A Legal And Policy Analysis Of
  Algorithmic Fairness}.
\newblock \bibinfo{journal}{\emph{Journal of Law, Technology \& Policy}}
  \bibinfo{volume}{20}, \bibinfo{number}{1} (\bibinfo{year}{2020}),
  \bibinfo{pages}{1--54}.
\newblock


\bibitem[Atrey(2019)]%
        {atrey2019intersectional}
\bibfield{author}{\bibinfo{person}{Shreya Atrey}.}
  \bibinfo{year}{2019}\natexlab{}.
\newblock \bibinfo{booktitle}{\emph{Intersectional discrimination}}.
\newblock \bibinfo{publisher}{Oxford University Press, USA}.
\newblock


\bibitem[Bell(2002)]%
        {bell2002anti}
\bibfield{author}{\bibinfo{person}{Mark Bell}.}
  \bibinfo{year}{2002}\natexlab{}.
\newblock \showarticletitle{Anti-discrimination law and the European Union}.
\newblock  (\bibinfo{year}{2002}).
\newblock


\bibitem[Bellamy et~al\mbox{.}(2018)]%
        {bellamy2018aif360}
\bibfield{author}{\bibinfo{person}{R.~K.~E. Bellamy}, \bibinfo{person}{K. Dey},
  \bibinfo{person}{M. Hind}, \bibinfo{person}{S.~C. Hoffman},
  \bibinfo{person}{S. Houde}, \bibinfo{person}{K. Kannan}, \bibinfo{person}{P.
  Lohia}, \bibinfo{person}{J. Martino}, \bibinfo{person}{S. Mehta},
  \bibinfo{person}{A. Mojsilovic}, \bibinfo{person}{S. Nagar},
  \bibinfo{person}{K.~Natesan Ramamurthy}, \bibinfo{person}{J. Richards},
  \bibinfo{person}{D. Saha}, \bibinfo{person}{P. Sattigeri},
  \bibinfo{person}{M. Singh}, \bibinfo{person}{K.~R. Varshney}, {and}
  \bibinfo{person}{Y. Zhang}.} \bibinfo{year}{2018}\natexlab{}.
\newblock \bibinfo{title}{{AI Fairness} 360: An Extensible Toolkit for
  Detecting, Understanding, and Mitigating Unwanted Algorithmic Bias}.
\newblock
\newblock
\urldef\tempurl%
\url{https://arxiv.org/abs/1810.01943}
\showURL{%
\tempurl}


\bibitem[Berk et~al\mbox{.}(2021)]%
        {berk2021fairness}
\bibfield{author}{\bibinfo{person}{Richard Berk}, \bibinfo{person}{Hoda
  Heidari}, \bibinfo{person}{Shahin Jabbari}, \bibinfo{person}{Michael Kearns},
  {and} \bibinfo{person}{Aaron Roth}.} \bibinfo{year}{2021}\natexlab{}.
\newblock \showarticletitle{Fairness in criminal justice risk assessments: The
  state of the art}.
\newblock \bibinfo{journal}{\emph{Sociological Methods \& Research}}
  \bibinfo{volume}{50}, \bibinfo{number}{1} (\bibinfo{year}{2021}),
  \bibinfo{pages}{3--44}.
\newblock


\bibitem[Biasiotti et~al\mbox{.}(2008)]%
        {biasiotti2008legal}
\bibfield{author}{\bibinfo{person}{Mariangela Biasiotti},
  \bibinfo{person}{Enrico Francesconi}, \bibinfo{person}{Monica Palmirani},
  \bibinfo{person}{Giovanni Sartor}, {and} \bibinfo{person}{Fabio Vitali}.}
  \bibinfo{year}{2008}\natexlab{}.
\newblock \showarticletitle{Legal informatics and management of legislative
  documents}.
\newblock \bibinfo{journal}{\emph{Global Center for ICT in Parliament Working
  Paper}}  \bibinfo{volume}{2} (\bibinfo{year}{2008}).
\newblock


\bibitem[Bullock and Masselot(2012)]%
        {bullock2012multiple}
\bibfield{author}{\bibinfo{person}{Jess Bullock} {and} \bibinfo{person}{Annick
  Masselot}.} \bibinfo{year}{2012}\natexlab{}.
\newblock \showarticletitle{Multiple discrimination and intersectional
  disadvantages: Challenges and opportunities in the European Union legal
  framework}.
\newblock \bibinfo{journal}{\emph{Colum. J. Eur. L.}}  \bibinfo{volume}{19}
  (\bibinfo{year}{2012}), \bibinfo{pages}{57}.
\newblock


\bibitem[Celis et~al\mbox{.}(2019)]%
        {celis2019classification}
\bibfield{author}{\bibinfo{person}{L~Elisa Celis}, \bibinfo{person}{Lingxiao
  Huang}, \bibinfo{person}{Vijay Keswani}, {and} \bibinfo{person}{Nisheeth~K
  Vishnoi}.} \bibinfo{year}{2019}\natexlab{}.
\newblock \showarticletitle{Classification with fairness constraints: A
  meta-algorithm with provable guarantees}. In
  \bibinfo{booktitle}{\emph{Proceedings of the conference on fairness,
  accountability, and transparency}}. \bibinfo{pages}{319--328}.
\newblock


\bibitem[Citron and Pasquale(2014)]%
        {citron2014scored}
\bibfield{author}{\bibinfo{person}{Danielle~Keats Citron} {and}
  \bibinfo{person}{Frank Pasquale}.} \bibinfo{year}{2014}\natexlab{}.
\newblock \showarticletitle{The scored society: Due process for automated
  predictions}.
\newblock \bibinfo{journal}{\emph{Wash. L. Rev.}}  \bibinfo{volume}{89}
  (\bibinfo{year}{2014}), \bibinfo{pages}{1}.
\newblock


\bibitem[Commission(2019)]%
        {HLEG}
\bibfield{author}{\bibinfo{person}{European Commission}.}
  \bibinfo{year}{2019}\natexlab{}.
\newblock \bibinfo{booktitle}{\emph{Ethic Guidelines for Trustworthy AI}}.
\newblock
\urldef\tempurl%
\url{https://ec.europa.eu/newsroom/dae/document.cfm?doc_id=60419}
\showURL{%
\tempurl}


\bibitem[d'Alessandro et~al\mbox{.}(2017)]%
        {dalessandro2017conscientious}
\bibfield{author}{\bibinfo{person}{Brian d'Alessandro}, \bibinfo{person}{Cathy
  O'Neil}, {and} \bibinfo{person}{Tom {LaGatta}}.}
  \bibinfo{year}{2017}\natexlab{}.
\newblock \showarticletitle{Conscientious Classification: A Data Scientist's
  Guide to Discrimination-Aware Classification}.
\newblock  \bibinfo{volume}{5}, \bibinfo{number}{2} (\bibinfo{year}{2017}),
  \bibinfo{pages}{120--134}.
\newblock


\bibitem[Datta et~al\mbox{.}(2015)]%
        {datta_automated_2015}
\bibfield{author}{\bibinfo{person}{Amit Datta}, \bibinfo{person}{Michael~Carl
  Tschantz}, {and} \bibinfo{person}{Anupam Datta}.}
  \bibinfo{year}{2015}\natexlab{}.
\newblock \showarticletitle{Automated Experiments on Ad Privacy Settings: A
  Tale of Opacity, Choice, and Discrimination}.
\newblock  \bibinfo{volume}{2015}, \bibinfo{number}{1} (\bibinfo{year}{2015}),
  \bibinfo{pages}{92--112}.
\newblock


\bibitem[De~Vos(2020)]%
        {devos}
\bibfield{author}{\bibinfo{person}{Marc De~Vos}.}
  \bibinfo{year}{2020}\natexlab{}.
\newblock \showarticletitle{The European Court of Justice and the march towards
  substantive equality in European Union anti-discrimination law}.
\newblock \bibinfo{journal}{\emph{International Journal of Discrimination and
  the Law}} \bibinfo{volume}{20}, \bibinfo{number}{1} (\bibinfo{year}{2020}),
  \bibinfo{pages}{62--87}.
\newblock


\bibitem[Dietz and Kleinlogel(2015)]%
        {dietz2015employment}
\bibfield{author}{\bibinfo{person}{Joerg Dietz} {and}
  \bibinfo{person}{Emmanuelle~P Kleinlogel}.} \bibinfo{year}{2015}\natexlab{}.
\newblock \showarticletitle{Employment discrimination as unethical behavior}.
\newblock In \bibinfo{booktitle}{\emph{The Oxford Handbook of Workplace
  Discrimination}}. \bibinfo{publisher}{Oxford University Press Oxford}.
\newblock


\bibitem[Dolata et~al\mbox{.}(2022)]%
        {zora207228}
\bibfield{author}{\bibinfo{person}{Mateusz Dolata}, \bibinfo{person}{Stefan
  Feuerriegel}, {and} \bibinfo{person}{Gerhard Schwabe}.}
  \bibinfo{year}{2022}\natexlab{}.
\newblock \showarticletitle{A sociotechnical view of algorithmic fairness}.
\newblock \bibinfo{journal}{\emph{Information Systems Journal}}
  \bibinfo{volume}{32}, \bibinfo{number}{4} (\bibinfo{date}{July}
  \bibinfo{year}{2022}), \bibinfo{pages}{754--818}.
\newblock
\showISSN{1350-1917}
\urldef\tempurl%
\url{https://doi.org/10.5167/uzh-207228}
\showURL{%
\tempurl}


\bibitem[Dressel and Farid(2018)]%
        {dressel2018compas}
\bibfield{author}{\bibinfo{person}{Julia Dressel} {and} \bibinfo{person}{Hany
  Farid}.} \bibinfo{year}{2018}\natexlab{}.
\newblock \showarticletitle{The accuracy, fairness, and limits of predicting
  recidivism}.
\newblock   \bibinfo{volume}{4} (\bibinfo{year}{2018}).
\newblock


\bibitem[Dua and Graff(2017)]%
        {Dua2019adult}
\bibfield{author}{\bibinfo{person}{Dheeru Dua} {and} \bibinfo{person}{Casey
  Graff}.} \bibinfo{year}{2017}\natexlab{}.
\newblock \bibinfo{title}{{UCI} Machine Learning Repository}.
\newblock
\newblock
\urldef\tempurl%
\url{http://archive.ics.uci.edu/ml}
\showURL{%
\tempurl}


\bibitem[Dwork et~al\mbox{.}(2012)]%
        {dwork2012fairness}
\bibfield{author}{\bibinfo{person}{Cynthia Dwork}, \bibinfo{person}{Moritz
  Hardt}, \bibinfo{person}{Toniann Pitassi}, \bibinfo{person}{Omer Reingold},
  {and} \bibinfo{person}{Richard Zemel}.} \bibinfo{year}{2012}\natexlab{}.
\newblock \showarticletitle{Fairness through awareness}. In
  \bibinfo{booktitle}{\emph{Proceedings of the 3rd Innovations in Theoretical
  Computer Science Conference}} (New York, {NY}, {USA})
  \emph{(\bibinfo{series}{{ITCS} '12})}. \bibinfo{publisher}{Association for
  Computing Machinery}, \bibinfo{pages}{214--226}.
\newblock


\bibitem[Ellis and Watson(2012)]%
        {ellis2012eu}
\bibfield{author}{\bibinfo{person}{Evelyn Ellis} {and}
  \bibinfo{person}{Philippa Watson}.} \bibinfo{year}{2012}\natexlab{}.
\newblock \bibinfo{booktitle}{\emph{EU anti-discrimination law}}.
\newblock \bibinfo{publisher}{OUP Oxford}.
\newblock


\bibitem[{European Commission}(2021)]%
        {AIA}
\bibfield{author}{\bibinfo{person}{{European Commission}}.}
  \bibinfo{year}{2021}\natexlab{}.
\newblock \bibinfo{title}{Proposal for a Regulation Of The European Parliament
  And Of The Council Laying Down Harmonised Rules On Artificial Intelligence
  (Artificial Intelligence Act) And Amending Certain Union Legislative Acts.
  COM/2021/206 final}.
\newblock
\newblock


\bibitem[Feldman et~al\mbox{.}(2015)]%
        {feldman2015certifying}
\bibfield{author}{\bibinfo{person}{Michael Feldman}, \bibinfo{person}{Sorelle~A
  Friedler}, \bibinfo{person}{John Moeller}, \bibinfo{person}{Carlos
  Scheidegger}, {and} \bibinfo{person}{Suresh Venkatasubramanian}.}
  \bibinfo{year}{2015}\natexlab{}.
\newblock \showarticletitle{Certifying and removing disparate impact}. In
  \bibinfo{booktitle}{\emph{proceedings of the 21th ACM SIGKDD international
  conference on knowledge discovery and data mining}}.
  \bibinfo{pages}{259--268}.
\newblock


\bibitem[Freedman et~al\mbox{.}(2007)]%
        {freedan2007}
\bibfield{author}{\bibinfo{person}{David Freedman}, \bibinfo{person}{Robert
  Pisani}, {and} \bibinfo{person}{Roger Purves}.}
  \bibinfo{year}{2007}\natexlab{}.
\newblock \bibinfo{booktitle}{\emph{Statistics}}.
\newblock \bibinfo{publisher}{W. W. Norton \& Company}, \bibinfo{address}{W. W.
  Norton \& Company}.
\newblock


\bibitem[Friedman and Nissenbaum(1996)]%
        {friedman1996bias}
\bibfield{author}{\bibinfo{person}{Batya Friedman} {and} \bibinfo{person}{Helen
  Nissenbaum}.} \bibinfo{year}{1996}\natexlab{}.
\newblock \showarticletitle{Bias in computer systems}.
\newblock \bibinfo{journal}{\emph{ACM Transactions on Information Systems
  (TOIS)}} \bibinfo{volume}{14}, \bibinfo{number}{3} (\bibinfo{year}{1996}),
  \bibinfo{pages}{330--347}.
\newblock


\bibitem[Hacker(2018)]%
        {hacker2018teaching}
\bibfield{author}{\bibinfo{person}{Philipp Hacker}.}
  \bibinfo{year}{2018}\natexlab{}.
\newblock \showarticletitle{Teaching fairness to artificial intelligence:
  existing and novel strategies against algorithmic discrimination under EU
  law}.
\newblock \bibinfo{journal}{\emph{Common Market Law Review}}
  \bibinfo{volume}{55}, \bibinfo{number}{4} (\bibinfo{year}{2018}).
\newblock


\bibitem[Helberger et~al\mbox{.}(2020)]%
        {helberger2020fairest}
\bibfield{author}{\bibinfo{person}{Natali Helberger}, \bibinfo{person}{Theo
  Araujo}, {and} \bibinfo{person}{Claes~H de Vreese}.}
  \bibinfo{year}{2020}\natexlab{}.
\newblock \showarticletitle{Who is the fairest of them all? Public attitudes
  and expectations regarding automated decision-making}.
\newblock \bibinfo{journal}{\emph{Computer Law \& Security Review}}
  \bibinfo{volume}{39} (\bibinfo{year}{2020}), \bibinfo{pages}{105456}.
\newblock


\bibitem[Hort et~al\mbox{.}(2022)]%
        {hort2022survey}
\bibfield{author}{\bibinfo{person}{Max Hort}, \bibinfo{person}{Zhenpeng Chen},
  \bibinfo{person}{J. Zhang}, \bibinfo{person}{Federica Sarro}, {and}
  \bibinfo{person}{M. Harman}.} \bibinfo{year}{2022}\natexlab{}.
\newblock \showarticletitle{Bias Mitigation for Machine Learning Classifiers: A
  Comprehensive Survey}.
\newblock \bibinfo{journal}{\emph{ArXiv}} (\bibinfo{year}{2022}).
\newblock


\bibitem[{Julia Angwin} et~al\mbox{.}(2016)]%
        {angwin2016propublica}
\bibfield{author}{\bibinfo{person}{{Julia Angwin}}, \bibinfo{person}{{Jeff
  Larson}}, \bibinfo{person}{{Surya Mattu}}, {and} \bibinfo{person}{{Lauren
  Kirchner}}.} \bibinfo{year}{2016}\natexlab{}.
\newblock \bibinfo{booktitle}{\emph{Machine Bias: There’s software used
  across the country to predict future criminals. And it’s biased against
  blacks.}}
\newblock
\urldef\tempurl%
\url{https://www.propublica.org/article/machine-bias-risk-assessments-in-criminal-sentencing}
\showURL{%
\tempurl}


\bibitem[Kamiran et~al\mbox{.}(2013)]%
        {kamiran2013quantifying}
\bibfield{author}{\bibinfo{person}{Faisal Kamiran}, \bibinfo{person}{Indr{\.e}
  {\v{Z}}liobait{\.e}}, {and} \bibinfo{person}{Toon Calders}.}
  \bibinfo{year}{2013}\natexlab{}.
\newblock \showarticletitle{Quantifying explainable discrimination and removing
  illegal discrimination in automated decision making}.
\newblock \bibinfo{journal}{\emph{Knowledge and information systems}}
  \bibinfo{volume}{35}, \bibinfo{number}{3} (\bibinfo{year}{2013}),
  \bibinfo{pages}{613--644}.
\newblock


\bibitem[Kelly(2021)]%
        {kelly2021tale}
\bibfield{author}{\bibinfo{person}{Anthony Kelly}.}
  \bibinfo{year}{2021}\natexlab{}.
\newblock \showarticletitle{A tale of two algorithms: The appeal and repeal of
  calculated grades systems in England and Ireland in 2020}.
\newblock \bibinfo{journal}{\emph{British Educational Research Journal}}
  \bibinfo{volume}{47}, \bibinfo{number}{3} (\bibinfo{year}{2021}),
  \bibinfo{pages}{725--741}.
\newblock


\bibitem[Kleinberg et~al\mbox{.}(2018)]%
        {kleinberg2018algorithmic}
\bibfield{author}{\bibinfo{person}{Jon Kleinberg}, \bibinfo{person}{Jens
  Ludwig}, \bibinfo{person}{Sendhil Mullainathan}, {and}
  \bibinfo{person}{Ashesh Rambachan}.} \bibinfo{year}{2018}\natexlab{}.
\newblock \showarticletitle{Algorithmic fairness}. In
  \bibinfo{booktitle}{\emph{Aea papers and proceedings}},
  Vol.~\bibinfo{volume}{108}. \bibinfo{pages}{22--27}.
\newblock


\bibitem[Kleinberg et~al\mbox{.}(2016)]%
        {kleinberg2016inherent}
\bibfield{author}{\bibinfo{person}{Jon Kleinberg}, \bibinfo{person}{Sendhil
  Mullainathan}, {and} \bibinfo{person}{Manish Raghavan}.}
  \bibinfo{year}{2016}\natexlab{}.
\newblock \showarticletitle{Inherent trade-offs in the fair determination of
  risk scores}.
\newblock \bibinfo{journal}{\emph{arXiv preprint arXiv:1609.05807}}
  (\bibinfo{year}{2016}).
\newblock


\bibitem[Lambrecht and Tucker(2018)]%
        {lambrecht_algorithmic_2018}
\bibfield{author}{\bibinfo{person}{Anja Lambrecht} {and}
  \bibinfo{person}{Catherine~E. Tucker}.} \bibinfo{year}{2018}\natexlab{}.
\newblock \bibinfo{title}{Algorithmic Bias? An Empirical Study into Apparent
  Gender-Based Discrimination in the Display of {STEM} Career Ads}.
\newblock
\newblock


\bibitem[Le~Quy et~al\mbox{.}(2022)]%
        {le_quy2022datasets}
\bibfield{author}{\bibinfo{person}{Tai Le~Quy}, \bibinfo{person}{Arjun Roy},
  \bibinfo{person}{Vasileios Iosifidis}, \bibinfo{person}{Wenbin Zhang}, {and}
  \bibinfo{person}{Eirini Ntoutsi}.} \bibinfo{year}{2022}\natexlab{}.
\newblock \showarticletitle{A survey on datasets for fairness-aware machine
  learning}.
\newblock  \bibinfo{volume}{12}, \bibinfo{number}{3} (\bibinfo{year}{2022}),
  \bibinfo{pages}{e1452}.
\newblock


\bibitem[MacKinnon(1991)]%
        {mackinnon1991reflections}
\bibfield{author}{\bibinfo{person}{Catharine~A MacKinnon}.}
  \bibinfo{year}{1991}\natexlab{}.
\newblock \showarticletitle{Reflections on sex equality under law}.
\newblock \bibinfo{journal}{\emph{Yale Law Journal}} (\bibinfo{year}{1991}),
  \bibinfo{pages}{1281--1328}.
\newblock


\bibitem[Madaio et~al\mbox{.}(2020)]%
        {madaio2020co}
\bibfield{author}{\bibinfo{person}{Michael~A Madaio}, \bibinfo{person}{Luke
  Stark}, \bibinfo{person}{Jennifer Wortman~Vaughan}, {and}
  \bibinfo{person}{Hanna Wallach}.} \bibinfo{year}{2020}\natexlab{}.
\newblock \showarticletitle{Co-designing checklists to understand
  organizational challenges and opportunities around fairness in AI}. In
  \bibinfo{booktitle}{\emph{Proceedings of the 2020 CHI Conference on Human
  Factors in Computing Systems}}. \bibinfo{pages}{1--14}.
\newblock


\bibitem[Makkonen(2007)]%
        {makkonen2007measuring}
\bibfield{author}{\bibinfo{person}{Timo Makkonen}.}
  \bibinfo{year}{2007}\natexlab{}.
\newblock \bibinfo{booktitle}{\emph{Measuring discrimination: Data collection
  and EU equality law}}.
\newblock \bibinfo{publisher}{Office for Official Publications of the European
  Communities}.
\newblock


\bibitem[Malgieri(2020)]%
        {malgieri2020concept}
\bibfield{author}{\bibinfo{person}{Gianclaudio Malgieri}.}
  \bibinfo{year}{2020}\natexlab{}.
\newblock \showarticletitle{The concept of fairness in the GDPR: a linguistic
  and contextual interpretation}. In \bibinfo{booktitle}{\emph{Proceedings of
  the 2020 Conference on fairness, accountability, and transparency}}.
  \bibinfo{pages}{154--166}.
\newblock


\bibitem[Manrai et~al\mbox{.}(2016)]%
        {manrai2016genetic}
\bibfield{author}{\bibinfo{person}{Arjun~K. Manrai}, \bibinfo{person}{Birgit~H.
  Funke}, \bibinfo{person}{Heidi~L. Rehm}, \bibinfo{person}{Morten~S. Olesen},
  \bibinfo{person}{Bradley~A. Maron}, \bibinfo{person}{Peter Szolovits},
  \bibinfo{person}{David~M. Margulies}, \bibinfo{person}{Joseph Loscalzo},
  {and} \bibinfo{person}{Isaac~S. Kohane}.} \bibinfo{year}{2016}\natexlab{}.
\newblock \showarticletitle{Genetic Misdiagnoses and the Potential for Health
  Disparities}.
\newblock  \bibinfo{volume}{375}, \bibinfo{number}{7} (\bibinfo{year}{2016}),
  \bibinfo{pages}{655--665}.
\newblock
\urldef\tempurl%
\url{https://doi.org/10.1056/NEJMsa1507092}
\showDOI{\tempurl}


\bibitem[Mehrabi et~al\mbox{.}(2021)]%
        {mehrabi2021survey}
\bibfield{author}{\bibinfo{person}{Ninareh Mehrabi}, \bibinfo{person}{Fred
  Morstatter}, \bibinfo{person}{Nripsuta Saxena}, \bibinfo{person}{Kristina
  Lerman}, {and} \bibinfo{person}{Aram Galstyan}.}
  \bibinfo{year}{2021}\natexlab{}.
\newblock \showarticletitle{A Survey on Bias and Fairness in Machine Learning}.
\newblock  \bibinfo{volume}{54}, \bibinfo{number}{6} (\bibinfo{year}{2021}),
  \bibinfo{pages}{115:1--115:35}.
\newblock


\bibitem[Mitchell et~al\mbox{.}(2021)]%
        {mitchell2021algorithmic}
\bibfield{author}{\bibinfo{person}{Shira Mitchell}, \bibinfo{person}{Eric
  Potash}, \bibinfo{person}{Solon Barocas}, \bibinfo{person}{Alexander
  D'Amour}, {and} \bibinfo{person}{Kristian Lum}.}
  \bibinfo{year}{2021}\natexlab{}.
\newblock \showarticletitle{Algorithmic fairness: Choices, assumptions, and
  definitions}.
\newblock \bibinfo{journal}{\emph{Annual Review of Statistics and Its
  Application}}  \bibinfo{volume}{8} (\bibinfo{year}{2021}),
  \bibinfo{pages}{141--163}.
\newblock


\bibitem[Moeckli et~al\mbox{.}(2010)]%
        {moeckli2010equality}
\bibfield{author}{\bibinfo{person}{Daniel Moeckli} {et~al\mbox{.}}}
  \bibinfo{year}{2010}\natexlab{}.
\newblock \showarticletitle{Equality and non-discrimination}.
\newblock \bibinfo{journal}{\emph{International human rights law}}
  (\bibinfo{year}{2010}), \bibinfo{pages}{189--208}.
\newblock


\bibitem[Roselli et~al\mbox{.}(2019)]%
        {roselli2019managing}
\bibfield{author}{\bibinfo{person}{Drew Roselli}, \bibinfo{person}{Jeanna
  Matthews}, {and} \bibinfo{person}{Nisha Talagala}.}
  \bibinfo{year}{2019}\natexlab{}.
\newblock \showarticletitle{Managing bias in AI}. In
  \bibinfo{booktitle}{\emph{Companion Proceedings of The 2019 World Wide Web
  Conference}}. \bibinfo{pages}{539--544}.
\newblock


\bibitem[Sapienza(2022)]%
        {sapienza2022p}
\bibfield{author}{\bibinfo{person}{Salvatore Sapienza}.}
  \bibinfo{year}{2022}\natexlab{}.
\newblock \showarticletitle{Group Pirvacy and Dietary Preferences}.
\newblock In \bibinfo{booktitle}{\emph{Big Data, Algorithms and Food Safety: A
  Legal and Ethical Approach to Data Ownership and Data Governance}}.
  \bibinfo{publisher}{Springer}, \bibinfo{pages}{106--108}.
\newblock


\bibitem[Schneider(2020)]%
        {housing}
\bibfield{author}{\bibinfo{person}{Valerie Schneider}.}
  \bibinfo{year}{2020}\natexlab{}.
\newblock \showarticletitle{Locked out by big data: how big data algorithms and
  machine learning may undermine housing justice}.
\newblock \bibinfo{journal}{\emph{Colum. Hum. Rts. L. Rev.}}
  \bibinfo{volume}{52} (\bibinfo{year}{2020}), \bibinfo{pages}{251}.
\newblock


\bibitem[SEAPHE(2007)]%
        {WinNT}
\bibfield{author}{\bibinfo{person}{Project SEAPHE}.}
  \bibinfo{year}{2007}\natexlab{}.
\newblock \bibinfo{booktitle}{\emph{{Project SEAPHE} Law School Admissions}}.
\newblock
\urldef\tempurl%
\url{http://www.seaphe.org/databases.php}
\showURL{%
\tempurl}


\bibitem[Siapka(2018)]%
        {siapka2018ethical}
\bibfield{author}{\bibinfo{person}{Anastasia Siapka}.}
  \bibinfo{year}{2018}\natexlab{}.
\newblock \showarticletitle{The Ethical and Legal Challenges of Artificial
  Intelligence: The EU response to biased and discriminatory AI}.
\newblock \bibinfo{journal}{\emph{Available at SSRN 3408773}}
  (\bibinfo{year}{2018}).
\newblock


\bibitem[Srivastava et~al\mbox{.}(2019)]%
        {srivastava2019mathematical}
\bibfield{author}{\bibinfo{person}{Megha Srivastava}, \bibinfo{person}{Hoda
  Heidari}, {and} \bibinfo{person}{Andreas Krause}.}
  \bibinfo{year}{2019}\natexlab{}.
\newblock \showarticletitle{Mathematical notions vs. human perception of
  fairness: A descriptive approach to fairness for machine learning}. In
  \bibinfo{booktitle}{\emph{Proceedings of the 25th ACM SIGKDD international
  conference on knowledge discovery \& data mining}}.
  \bibinfo{pages}{2459--2468}.
\newblock


\bibitem[Verma and Rubin(2018)]%
        {verma2018fairness}
\bibfield{author}{\bibinfo{person}{Sahil Verma} {and} \bibinfo{person}{Julia
  Rubin}.} \bibinfo{year}{2018}\natexlab{}.
\newblock \showarticletitle{Fairness definitions explained}. In
  \bibinfo{booktitle}{\emph{2018 ieee/acm international workshop on software
  fairness (fairware)}}. IEEE, \bibinfo{publisher}{ACM}, \bibinfo{pages}{1--7}.
\newblock


\bibitem[Wachter(2022)]%
        {wachter2022theory}
\bibfield{author}{\bibinfo{person}{Sandra Wachter}.}
  \bibinfo{year}{2022}\natexlab{}.
\newblock \showarticletitle{The theory of artificial immutability: Protecting
  algorithmic groups under anti-discrimination law}.
\newblock \bibinfo{journal}{\emph{arXiv preprint arXiv:2205.01166}}
  (\bibinfo{year}{2022}).
\newblock


\bibitem[Wachter et~al\mbox{.}(2020)]%
        {wachter2020bias}
\bibfield{author}{\bibinfo{person}{Sandra Wachter}, \bibinfo{person}{Brent
  Mittelstadt}, {and} \bibinfo{person}{Chris Russell}.}
  \bibinfo{year}{2020}\natexlab{}.
\newblock \showarticletitle{Bias preservation in machine learning: the legality
  of fairness metrics under EU non-discrimination law}.
\newblock \bibinfo{journal}{\emph{W. Va. L. Rev.}}  \bibinfo{volume}{123}
  (\bibinfo{year}{2020}), \bibinfo{pages}{735}.
\newblock


\bibitem[Wachter et~al\mbox{.}(2021)]%
        {wachter2021fairness}
\bibfield{author}{\bibinfo{person}{Sandra Wachter}, \bibinfo{person}{Brent
  Mittelstadt}, {and} \bibinfo{person}{Chris Russell}.}
  \bibinfo{year}{2021}\natexlab{}.
\newblock \showarticletitle{Why fairness cannot be automated: Bridging the gap
  between EU non-discrimination law and AI}.
\newblock \bibinfo{journal}{\emph{Computer Law \& Security Review}}
  \bibinfo{volume}{41} (\bibinfo{year}{2021}), \bibinfo{pages}{105567}.
\newblock


\bibitem[Yousefi(2022)]%
        {yousefi2022notions}
\bibfield{author}{\bibinfo{person}{Yasaman Yousefi}.}
  \bibinfo{year}{2022}\natexlab{}.
\newblock \showarticletitle{Notions of Fairness in Automated Decision Making:
  An Interdisciplinary Approach to Open Issues}. In
  \bibinfo{booktitle}{\emph{Electronic Government and the Information Systems
  Perspective: 11th International Conference, EGOVIS 2022, August 22--24, 2022,
  Proceedings}}. \bibinfo{publisher}{Springer}, \bibinfo{address}{Vienna,
  Austria}, \bibinfo{pages}{3--17}.
\newblock


\bibitem[Zehlike et~al\mbox{.}(2020)]%
        {zehlike2020}
\bibfield{author}{\bibinfo{person}{Meike Zehlike}, \bibinfo{person}{Philipp
  Hacker}, {and} \bibinfo{person}{Emil Wiedemann}.}
  \bibinfo{year}{2020}\natexlab{}.
\newblock \showarticletitle{Matching code and law\: achieving algorithmic
  fairness with optimal transport}.
\newblock \bibinfo{journal}{\emph{Data Mining and Knowledge Discovery}}
  \bibinfo{volume}{34} (\bibinfo{year}{2020}), \bibinfo{pages}{163–200}.
\newblock


\end{thebibliography}

    \end{document}